\documentclass[11pt]{article}
\usepackage[dvipdfmx]{graphicx}
\usepackage{color}
\begin{document}

\begin{center}{\Large Information Criterion for the Gaussian and/or Laplace Distribution Models}\\

\vspace{8mm}
{\large Genshiro Kitagawa}\\[1mm]
The Institute of Statistical Mathematics\\[-1mm]
and\\[-1mm]
Graduate University for Advanced Study\\[2mm]

\vspace{5mm}
{\today}

\end{center}

\noindent{\bf Abstract}

The information criterion AIC has been used successfully in many areas of statistical modeling, and since it is derived based on the Taylor expansion of the log-likelihood function and the asymptotic distribution of the maximum likelihood estimator, it is not directly justified for likelihood functions that include non-differentiable points such as the Laplace distribution. In fact, it is known to work effectively in many such cases. In this paper, we attempt to evaluate the bias correction directly for the case where the true model or the model to be estimated is a simple Laplace distribution model. As a result, an approximate expression for the bias correction term was obtained. Numerical results show that the AIC approximations are relatively good except when the Gauss distribution model is fitted to data following the Laplace distribution.

\vspace{1mm}
\noindent

\section{Brief Review of Information Criteria for Model Evaluation}

\subsection{Evaluation of Statistical Models}
Given true distribution $g(x)$ and a statistical model $f(x|\theta)$ 
where $\theta$ is the parameter of the model, 
the Kullback-Leibler divergence (or Kullback-Leibler information)
is defined by
\begin{eqnarray}
  D_{KL}(g;f) &=& E_G[\log f(X|\mu,\sigma)] - E_G[\log f(X|\xi,\tau)],
\end{eqnarray}
where $E_G$ denotes the expectation with respect to the true distribution $g(x)$.
By calculating this Kullback-Leibler Divergence, the adequacy of the model $f(x)$ can be evaluated. However, in actual situations of statistical modeling, this divergence cannot be calculated directly because the true distribution is unknown.

\subsection{Asymptotic Unbiased Estimation of Expected Log-likelihood}
The log-likelihood is a natural estimator of the expected log-likelihood and the maximum likelihood method can be used for estimation of the parameters of the model.
If there are several candidate parametric models, it seems natural to estimate the  parameters by the maximum likelihood method, and then find the best model by comparing the values of the maximum log-likelihood \( \ell( \hat \theta ) \). However, in actuality, the maximum log-likelihood is not directly available for comparisons among several parametric models whose parameters are estimated by the maximum likelihood method, because of the presence of the bias.
That is, for the model with the maximum likelihood estimate \( \hat \theta \), the maximum log-likelihood, $ \ell( \hat\theta)$ has a positive bias as an estimator of $N{\rm E}_Y[ \log f(Y|\hat \theta)]$ (Akaike (1973,1974), Sakamoto et al. (1986), Konishi and Kitagawa (2008)).

This bias is caused by using the same data twice for the estimation of the parameters of the model and for the estimation of the expected log-likelihood for evaluation of the model.
The bias of \( \ell (\hat \theta ) \equiv \sum_{n=1}^N \log f(y_n|\hat \theta) \) as an estimate of \( N{\rm E}_Y [\log f(Y|\hat \theta )] \) is given by
\begin{equation}
 C \equiv {\rm E}_X\biggl[ N{\rm E}_Y[ \log f(Y|\hat \theta ) ]
-\sum_{n=1}^N \log f(y_n|\hat \theta ) \biggr]. \label{Eq_bias of log-likelihood}
\end{equation}
Note here that the maximum likelihood estimate $\hat\theta$ depends on the sample $X$ and can be expressed as $\hat\theta (X)$, and the expectation ${\rm E}_X$ is taken with respect to the true distribution of $X$.

Then, correcting the maximum log-likelihood $\ell (\hat\theta )$ for the bias $C$, $ \ell (\hat\theta )+C$ becomes an unbiased estimate of the expected log-likelihood \( N E_Y [\log f(Y|\hat \theta )] \). Here, as will be shown later, since the bias is evaluated as $C=-k$, we obtain the Akaike Information Criterion (AIC)\index{Akaike information criterion}\index{AIC} (Akaike (1973, 1974)).
\begin{eqnarray}
 {\rm AIC} = -2 \:\ell ( \hat \theta ) + 2\: k , \label{AIC}
\end{eqnarray}
where $k$ is the number of parameters.

\subsection{TIC: Takeuchi's Information Criterion}

\begin{figure}[h]
\begin{center}
\includegraphics[width=78mm,angle=0,clip=]{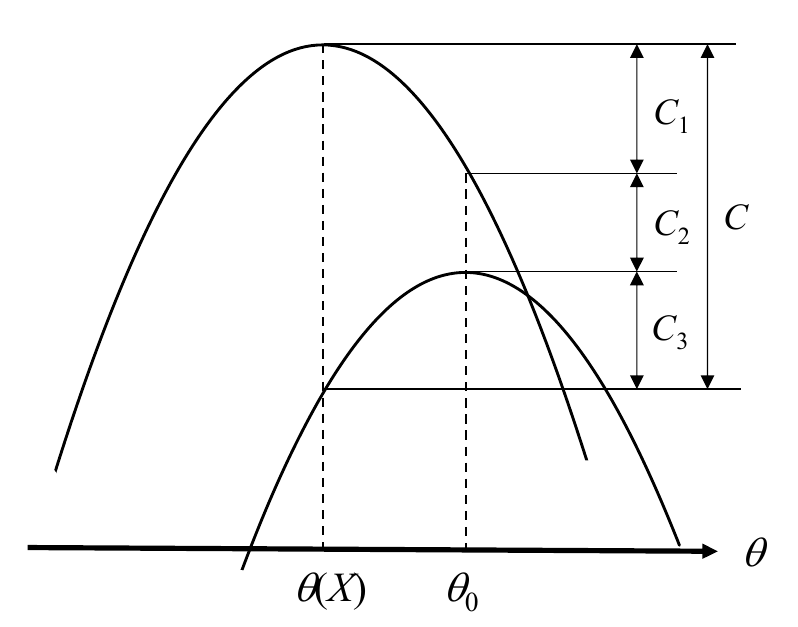}
\includegraphics[width=75mm,angle=0,clip=]{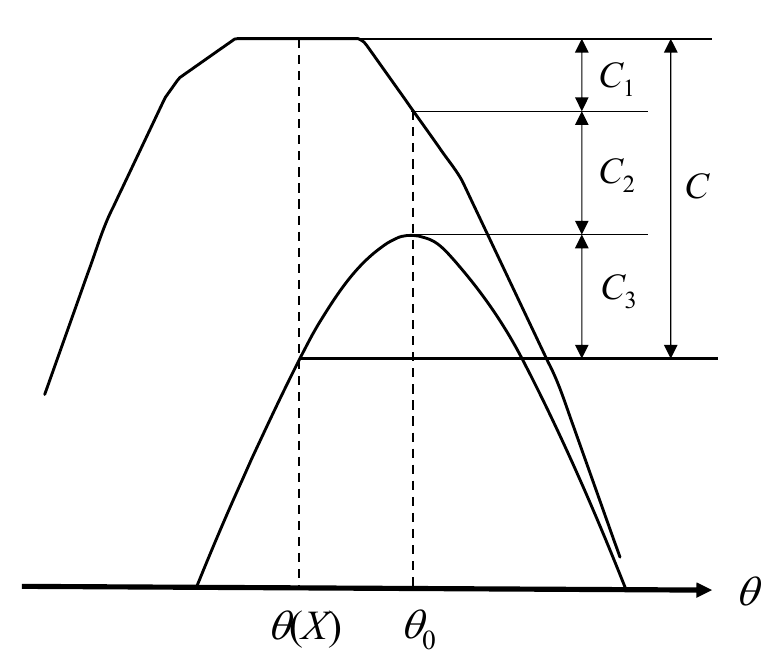}
\end{center}
\caption{ Difference between the expected log-likelihood and the log-likelihood and its decomposition into three terms. Left plot: Gaussian distribution model, right plot: Laplace distribution model. }
\label{Fig_bias_AIC}
\end{figure}

Here is a brief introduction to the derivation of AIC by Takeuchi (1976) (see also Konishi and Kitagawa (2008)).
It is assumed that the true distribution is $g(y)$, the model distribution is $f(y|\theta)$ and the maximum likelihood estimate of the parameter \( \theta \) based on data \(X = (x_1,\cdots,x_N)\) is denoted by \( \hat \theta \equiv \hat \theta (X )\). On the other hand, the parameter $\theta_0$ that maximizes the expected log-likelihood $ {\rm E}_ Y [\log f (Y|\theta)] \) is called the true parameter. Then, $\theta_0$ satisfies
\[
\frac{ \partial}{\partial \theta} {\rm E}_Y [\log f(Y| \theta_0 )] = 0.
\]
On the other hand, since \( \hat \theta \) maximizes the log-likelihood function \( \ell(\theta) = \sum_{n=1}^{N} \log f(x_n|\theta)\), the following equation holds:
\[
\frac{\partial}{\partial \theta} \sum_{n=1}^{N} \log f(x_n|\hat{\theta}) = 0.
\]
Here, the terms in (\ref{Eq_bias of log-likelihood}) can be decomposed into three terms (see Figure \ref{Fig_bias_AIC}).
\begin{eqnarray}
 C &=& N {\rm E}_X \Bigl[ {\rm E}_Y [\log f(Y|\hat\theta )] 
-{\rm E}_Y [\log f(Y|\theta_0)]  \Bigr] \nonumber \\
 & & + \:\: N {\rm E}_X \Bigl[ {\rm E}_Y [\log f(Y|\theta_0 ) ]
-N^{-1}\sum_{n=1}^N \log f(x_n|\theta_0)\Bigr] \nonumber \\[-2mm]
 & & + \:\: N {\rm E}_X \Bigl[ N^{-1}\sum_{n=1}^N \log f(x_n|\theta_0) 
-N^{-1}\sum_{n=1}^N \log f(x_n|\hat\theta)\Bigr] \nonumber\\
 & \equiv& C_1 + C_2 + C_3. \label{Eq_decompsition_of_bias}
\end{eqnarray}
\begin{eqnarray}
C_1 &\approx& -\frac{1}{2}(\hat \theta -\theta_0)^T J ( \hat \theta -\theta_0) 
=\: {\rm trace}\left\{IJ^{-1}\right\} 
 \:\approx\: -\frac{k}{2} \nonumber \\
 C_2 &=&  0   \label{TIC_C123}\\
C_3 &\approx&  -\frac{1}{2}(\hat \theta -\theta_0)^T J ( \hat \theta -\theta_0) 
=\: {\rm trace}\left\{IJ^{-1}\right\} 
-\frac{k}{2}. \nonumber
\end{eqnarray}
where the Fisher information matrix $I(\theta_0)$ and the negative of expected Hssian $J(\theta_0)$  are defined by
\begin{eqnarray}
  I(\theta_0) &=& \left[ \begin{array}{ccc} 
     \frac{\partial \log f}{\partial\theta_1} \frac{\partial \log f}{\partial\theta_1} & \cdots & \frac{\partial \log f)}{\partial\theta_1} \frac{\partial \log f}{\partial\theta_k} \\
     \vdots & \ddots & \vdots \\
     \frac{\partial \log f}{\partial\theta_k} \frac{\partial \log f}{\partial\theta_1} & \cdots & \frac{\partial \log f}{\partial\theta_k} \frac{\partial \log f}{\partial\theta_k} \\
\end{array}\right] \\
  J(\theta_0) &=& - \left[ \begin{array}{ccc} 
     \frac{\partial^2 \log f(x|\theta)}{\partial\theta_1\partial\theta_1} & \cdots & \frac{\partial^2 \log f(x|\theta)}{\partial\theta_1\partial\theta_k} \\
     \vdots & \ddots & \vdots \\
     \frac{\partial^2 \log f(x|\theta)}{\partial\theta_k\partial\theta_1} & \cdots & \frac{\partial^2 \log f(x|\theta)}{\partial\theta_k\partial\theta_k} \\
\end{array}\right] .
\end{eqnarray}

Then an estimate of the bias correction term is obtained by (Takeuchi (1976), Konishi and Kitagawa (2008))
\begin{equation}
C_{\textrm{TIC}} = I(\theta_0) J(\theta_0)^{-1}, \nonumber
\end{equation}

Figure \ref{Fig_bias_AIC} shows the structure of the bias term shown in equation
(\ref{Eq_decompsition_of_bias}). 
In this decomposition, the expected value of $C_2$ is strictly zero but has variance proportional to the number of data (Konishi and Kitagawa (2008), Table \ref{Tab_Decomposition_of_C}). This fact is particularly important when determining the true value of the bias and in deriving the EIC based on the bootstrap method.

The left panel of Figure \ref{Fig_bias_AIC} shows the Gauss distribution model, in which the log likelihood is a quadratic function. On the other hand, the right figure is for the Laplace distribution, where the upper curve is the log likelihood and the lower curve is the mean log likelihood. To illustrate the problem clearly, the case with a very small number of data ($N$=8) is shown. The log-likelihood is a continuous function, but it is not differentiable at the data points, and furthermore, when the number of data is even, it is constant around the maximum value of the log-likelihood. As will be shown in a later section, this results in a Hessian matrix that is zero and the TIC cannot be defined.

\subsection{Akaike Information Criterion, AIC, and its Finite Correction}
 
By summing up the three expressions $(\ref{TIC_C123})$, we have the approximation 
\begin{equation}
C = {\rm E}_X \biggl[ N{\rm E}_Y \log f(Y|\hat \theta ) 
- \sum_{n=1}^N \log f(x_n|\hat \theta )\biggr] \approx 
-k, \nonumber
\end{equation}
and this shows that \( \ell (\hat \theta ) \) is larger than \(N {\rm E}_Y \log f (Y|\hat\theta)  \) by $k$ on average.

Therefore, it can be seen that 
\begin{equation}
 \ell (\hat\theta ) + C \: \approx \:  \ell (\hat \theta ) - k 	\end{equation}
is an approximately unbiased estimator of the expected log-likelihood \(N{\rm E}_Y \log f (Y|\hat \theta) \) of the maximum likelihood model. The Akaike information criterion (AIC)\index{Akaike information criterion}\index{AIC} is defined by multiplying (4.46) by \(-2\), i.e.,
\begin{eqnarray}
 {\rm AIC} = -2 \:\ell ( \hat \theta )\hspace{0.1cm}+\hspace{0.1cm} 2\: k. \label{AIC2}
\end{eqnarray}
Since minimizing the AIC is approximately equivalent to minimizing the K-L information, an approximately optimal model is obtained by selecting the model that minimizes AIC. A reasonable and automatic model selection thus becomes possible by using the AIC.

In the framework of the Gaussian regression model with $k$ regression coefficients, $a_1,\ldots,a_k$, a finite modification of the AIC is presented by Sugiura (1978).
In the evaluation of the bias term
\begin{eqnarray}
E_G\left[ \ell(\hat\theta) - NE_G[\log f(X|\hat\theta)] \right]
&=& E_G\left[ \frac{1}{2\hat\sigma^2}\left\{ \sigma^2 + \sum_{i=1}^k (\hat{a}_i -a_i)^2\right\}\right],
\end{eqnarray}
where $\sigma^2$ follows a $\chi^2$ distribution with $n-k$ degrees of freedom, and the expectation of its inverse is $n(n-k-2)^{-1}$, the finite correction of AIC is given by 
\begin{eqnarray}
C= \frac{2nk}{n-k-2}.
\end{eqnarray}
This correction term is asymptotically the same as the one of AIC, but for small samples it is larger than $2k$.
The advantage of finite corrections for AIC and AIC is that the amount of bias correction depends only on the number of data and does not require complex calculations.

\subsection{EIC: Bootstrap Information Criterion}
Bootstrap estimate of the bias correction term can be evaluated by
\begin{equation}
C^{\ast} = \frac{1}{NB} \sum_{i=1}^{NB} \left\{ \ell(X_N^{\ast}(i)|\hat{\theta}(X_N^{\ast}(i))\, -\, 
                                                \ell(X_N|\hat{\theta}(X_N^{\ast}(i)) \right\}, \nonumber
\end{equation}
where $NB$ is the number of bootstrap resampling, $X_N^{\ast}(i)$ is the $i$-th bootstrap sample with sample size $N$ and $\hat\theta(X_N^{\ast}))$ is the maximum likelihood estimate of the parameter based of the bootstrap sample.
However, this bias correction term contains large variance and is not reliable especially for large 
sample size $N$. The reason for this is that $C_2$ in Figure 1 has a mean of 0, but the variance increases in proportion to the number of data $N$.
Therefore, $C_1 + C_3$, which is $C_2$ removed from $C$, has the same mean as $C$ and a smaller variance (Table \ref{Tab_Decomposition_of_C}),
\begin{eqnarray}
C^{\ast\ast} &=& \frac{1}{NB} \sum_{i=1}^{NB} \left( D_1^{\ast} + D_3^{\ast}  \right) \nonumber \\
 &=& \frac{1}{NB} \sum_{i=1}^{NB} \left\{ 
     \ell(X_N^{\ast}(i)|\hat{\theta}(X_N^{\ast}(i))\, -\, \ell(X_N|\hat{\theta}(X_N^{\ast}(i)) \right. \nonumber \\
 && {}\hspace{10.5mm}\left.+ \,\,\ell(X_N^{\ast}(i)|\hat{\theta}(X_N^{\ast}(i))\, -\, \ell(X_N|\hat{\theta}(X_N^{\ast}(i)) 
\right\}. \nonumber
\end{eqnarray}
%

\begin{figure}[h]
\begin{center}
\includegraphics[width=80mm,angle=00,clip=]{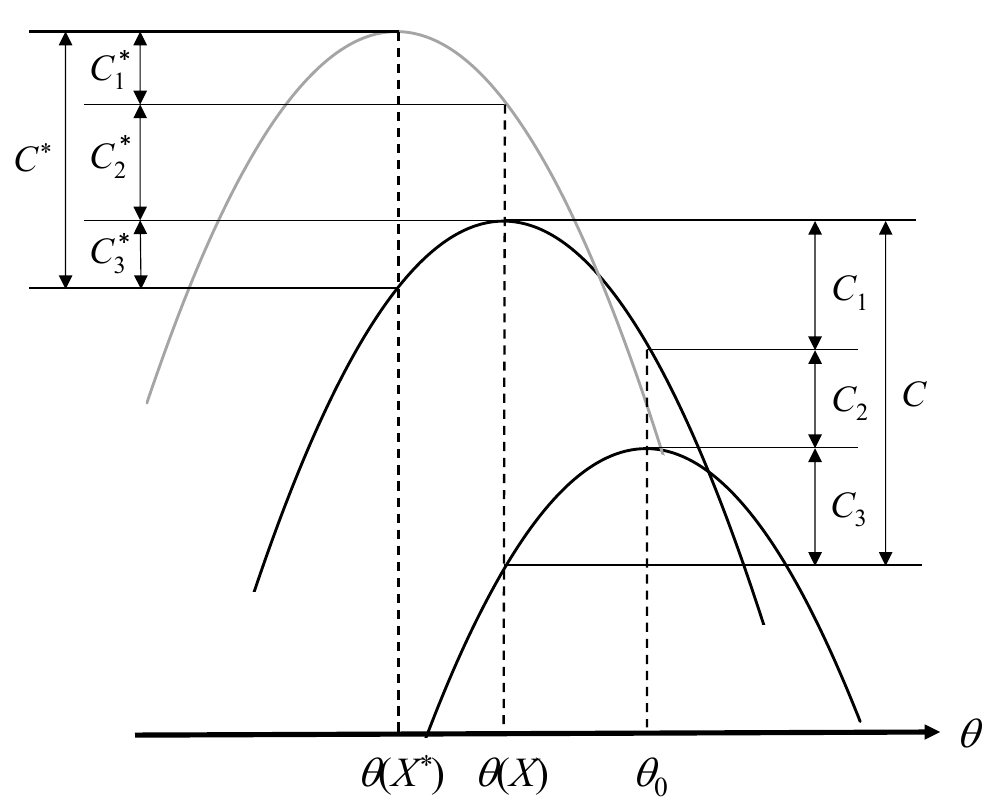}
\end{center}
\caption{ Difference between the expected log-likelihood and the log-likelihood. }
\label{Fig_bias_EIC}
\end{figure}

\section{Gaussian Distribution Model}

In this section, we consider a Gaussian distribution model, $N(\xi,\tau^2)$;
\begin{eqnarray}
  f(x|\xi ,\tau^2  ) &=& \frac{1}{\sqrt{2\pi\tau^2}} \exp\left\{ -\frac{(x-\xi)^2}{2\tau^2} \right\}.
\end{eqnarray}
The parameters of the model, $\xi$ and $\sigma^2$ are assumed to be unknown.
In the following two subsections, we consider the case where the true model generating the data is a Gaussian distribution and the Laplace distribution, respectively.

Given $N$ independent observations, $x_1,\ldots ,x_N$, the log-likelihood of the Gaussian model, $\ell (\xi, \tau^2 )$, is defined by
\begin{eqnarray}
  \ell (\xi, \tau^2 ) &=& -\frac{N}{2}\log 2\pi\tau^2 - \frac{1}{2\tau^2} \sum_{n=1}^N (x_n-\xi )^2.
\end{eqnarray}
Then the maximum likelihood estimates of the Gaussian distribution model are given by the sample mean and the sample variance defined by,
\begin{eqnarray}
   \hat{\xi} = \frac{1}{N}\sum_{n=1}^N x_n,\qquad    
   \hat{\tau}^2 = \frac{1}{N} \sum_{n=1}^N ( x_n -\hat{\xi} )^2. \nonumber
\end{eqnarray}
Substituting these estimates, the maximum log-likelihood is obtained by
\begin{eqnarray}
\ell(\hat{\xi},\hat{\tau}^2) &=& -\frac{N}{2}\log 2\pi\hat{\tau}^2 - \frac{N}{2}. 
\end{eqnarray}

For the Gaussian distribution model, from equations (\ref{Eq_Fisher-Information_Gauss}) and 
(\ref{Eq_Hessian_Gauss}), the bias estimates $ \textrm{tr}\left\{ I(\theta_0)J(\theta_0)^{-1} \right\} $ is given by
\begin{eqnarray}
I(\theta_0)J(\theta_0)^{-1} 
  &=& \left[ \begin{array}{cc} \frac{1}{\tau^2} & \frac{\mu_3}{2\tau^6} \\
           \frac{\mu_3}{2\tau^6} & \frac{\mu_4}{4\tau^8}-\frac{1}{4\tau^4}\end{array}\right]
      \left[ \begin{array}{cc} \tau^2 & 0 \\
           0 & 2\tau^4 \end{array} \right]
 = \left[ \begin{array}{cc} 1 & \frac{\mu_3}{\tau^2} \\
           \frac{\mu_3}{2\tau^4} & \frac{\mu_4}{2\tau^4}-\frac{1}{2}\end{array}\right]
\nonumber \\
  \textrm{tr}\left\{ I(\theta_0)J(\theta_0)^{-1} \right\} 
     &=& \frac{1}{2}\left( 1 + \frac{\mu_4}{\tau^4} \right),
\end{eqnarray}
where $\mu_4$ is the fourth-order central moment of the true distribution generating the data.

\subsection{For Gaussian data}
In this subsection, we consider the standard case that the true model generating the data is Gaussian distribution
with mean $\mu$ and the variance $\sigma^2$
\begin{eqnarray}
  g(x|\mu ,\sigma^2) &=& \frac{1}{\sqrt{2\pi\sigma^2}} \exp\left\{ -\frac{(x-\mu)^2}{2\sigma^2} \right\}.
\end{eqnarray}
Namely, in this case our model class contains the ^^ ^^ true model".
Given the model $f(x|\xi,\tau^2)$, the expected log-likelihood is given by
\begin{eqnarray}
E_G\bigl[ \log f(X|\xi,\tau^2)\big]  &=& -\frac{1}{2}\log 2\pi\tau^2 
 -\frac{1}{2\tau^2} E_G\left[ (x-\xi)^2 \right] \nonumber \\
&=& -\frac{1}{2}\log 2\pi\tau^2 
 -\frac{1}{2\tau^2} \left\{ \sigma^2 + (\mu -\xi)^2 \right\}, \nonumber
\end{eqnarray}
where $E_G$ denotes the expectation with respect to the Gaussian distribution $g(x|\mu,\sigma^2)$.

For the Gaussian distribution, since the fourth-order moment is given by $\mu_4 = 3\tau^4$,
the bias correction term is obtained by
\begin{eqnarray}
  \textrm{tr}\left\{ I(\theta_0)J(\theta_0)^{-1} \right\} 
     &=& \frac{1}{2}\left( 1 + \frac{\mu_4}{\tau^4} \right)
      =  \frac{1}{2}\biggl( 1 + \frac{3\tau^4}{\tau^4} \biggr) = 2.
\end{eqnarray}

The large-sample distributions of the mean and variance of the Gaussian model are given by
\begin{eqnarray}
  \hat{\mu} \sim N\biggl(\mu,\frac{\sigma^2}{N}\biggr) , \qquad
  \frac{N\hat{\sigma}^2}{\sigma^2} \sim \chi_N^2. \nonumber
\end{eqnarray}
Therefore, given the independence of the mean and the variance and the fact that the variance follows a $\chi^2_{N-1}$ distribution, the bias of the log-likelihood
as an estimate of the $N$ times of the expected log-likelihood is evaluated as
\begin{eqnarray}
\lefteqn{C = E_G\left[ \ell (\hat{\xi},\hat{\tau}^2) -N E_G[\log f(x|\hat{\xi},\hat{\tau}^2)] \right]}
\nonumber \\
 &=& E_G\left[ \frac{N}{2\hat{\tau}^2} \left\{ \sigma^2 + (\mu -\hat{\xi})^2 \right\} -\frac{N}{2}\right] \nonumber \\
 &=& \frac{N}{2}E_G\left[ \frac{\sigma^2}{\hat{\sigma}^2} \right]
        \left\{ 1 + \frac{1}{\sigma^2}E_G\left[ (\mu - \hat{\mu})^2 \right] \right\} - \frac{N}{2} \nonumber \\
 &=& \frac{N}{2}\frac{N}{N-3} \left( 1 + \frac{1}{N} \right) -\frac{N}{2} 
 = \frac{N}{2}\left( \frac{N}{N-3} + \frac{1}{N-3} -1 \right) = \frac{2N}{N-3}. \label{Eq_Bias_C_Gauss-Gauss}
\end{eqnarray}

\subsection{For Laplace Distribution Data}
In this subsection, we assume that the true distribution is the Laplace distribution defined by
\begin{eqnarray}
  g(x|\xi,\tau) = \frac{1}{2\tau}\exp\left(-\frac{|x-\xi|}{\tau}\right),
\end{eqnarray}
where $\xi$ is the location parameter and $\tau$ is the scale parameter.
The logarithm of the Gaussian distribution model and the expected value with respect to the true distribution is defined by (equation (\ref{Eq_expectation_of_x^2_Laplace}))
\begin{eqnarray}
  \log f(x|\xi ,\tau^2 ) &=& -\frac{1}{2}\log 2\pi\tau^2 - \frac{(x-\xi)^2}{2\tau^2} \nonumber \\
  E_L\left[ \log f(x|\xi ,\tau^2 ) \right] 
    &=& -\frac{1}{2}\log 2\pi\tau^2 - \frac{1}{2\tau^2}E_L\left[\, (x-\xi )^2\,\right] \nonumber \\ 
    &=& -\frac{1}{2}\log 2\pi\tau^2 - \frac{1}{2\tau^2}\left\{(\mu-\xi)^2 + 2\sigma^2 \right\}.   \nonumber 
\end{eqnarray}
The large-sample distributions of the maximum likelihood estimators of the  
mean and variance of the Gaussian model are given by
\begin{eqnarray}
  \hat{\xi} &\sim& N\left( \mu, \frac{2\sigma^2}{N} \right) \nonumber \\
  \frac{N\hat{\tau}^2}{2\sigma^2} &\sim&  \frac{\chi_N^2}{N},\qquad
     E[\hat{\tau}^2 ] = 2\sigma^2\frac{(N-2)}{N},\quad \textrm{Var}(\hat{\tau}^2) = \frac{8\sigma^4}{N}. \nonumber 
\end{eqnarray}
Here, using the asymptotic independence of the mean and the variance and 
equations (\ref{Eq_Laplace-Gauss_E[1/tau2]}), (\ref{Eq_expectation_of_x^2_Laplace}) and (\ref{Eq_Laplace-Gauss_Bias}), the bias of the log-likelihood
as an estimate of the $N$ times of the expected log-likelihood is evaluated as
\begin{eqnarray}
\lefteqn{C_N = E_L\left[ \ell (\hat{\xi},\hat{\tau}^2) -N E_L[\log f(x|\hat{\xi},\hat{\tau}^2)] \right]}
\nonumber \\
 &=& \frac{N}{2}E_L\left[ \frac{2\sigma^2 + (\hat{\xi}-\mu)^2}{\hat{\tau}^2} - 1 \right]
  =  \frac{N}{2}E_L\left[ \frac{\tau^2}{\hat{\tau}^2}\left\{ 1 + \frac{(\hat{\xi}-\mu)^2}{\tau^2}\right\} - 1 \right] \nonumber \\
 &=& \frac{N}{2}E_L\left[ \frac{\tau^2}{\hat{\tau}^2} \right]
        \left\{ 1 + \frac{1}{\tau^2} E_L\left[ (\hat{\xi}-\mu)^2 \right] \right\} - \frac{N}{2} \nonumber \\
 &\approx& \frac{N}{2}\left( 1 + \frac{6}{N} + \frac{21}{N^2} + \frac{46}{N^3}\right)
      \left( 1 + \frac{2\sigma^2}{N\tau^2} \right) -\frac{N}{2}   \nonumber \\
 &\approx&  \frac{N}{2}\left( 1 + \frac{7}{N} + \frac{27}{N^2} + \frac{67}{N^3} -1 \right)  
  \approx   \frac{7}{2} + \frac{27}{2N}. \label{Eq_Bias_C_Gauss-Laplace} 
\end{eqnarray}

For the Laplacian distribution, since the fourth order moment is given by $\mu_4 = 24\sigma^4=6\tau^4$,
the bias correction term is obtained by
\begin{eqnarray}
  \textrm{tr}\left\{ I(\theta_0)J(\theta_0)^{-1} \right\} 
     &=& \frac{1}{2}\left( 1 + \frac{\mu_4}{\tau^4} \right)
      =  \frac{1}{2}\left( 1 + \frac{6\tau^4}{\tau^4} \right) = 3.5.
\end{eqnarray}

\subsection{Numerical Example}
Table {\ref{Tab_Gaussian_model} shows the bias correction term evaluated by various methods.
The left and right halves of the table show the results when the true model generating the data is the Gaussian and Laplace distribution, respectively.
^^ ^^ True" evaluates the difference between the mean log-likelihood and the maximum log-likelihood of the model determined by the maximum likelihood estimates of the parameters using the $10^8$ Monte Carlo experiments for $n$=25 and 100, $10^7$ for $n$=400, and $10^6$ for $n$=1,600. 
In addition, in evaluating the bias, the variance reduction method is applied by using the decomposition in (4), i.e., we evaluate $C_1 + C_3$ instaed of $C$.
Since the number of parameters is 2, the correction term for AIC is always 2. $IJ^{-1}$ is the correction term for TIC. Although this is theoretically appealing, it contains an unknown variance and fourth-order moment $\mu_4$, which must be estimated from the data for actual use, and the variance of the bias correction term may be large. $C_n$ is the approximation of the direct evaluation of bias obtained in (\ref{Eq_Bias_C_Gauss-Gauss}) or (\ref{Eq_Bias_C_Gauss-Laplace}). $B_n$ is the correction term estimated by bootstrap method.

\begin{table}[h]
\caption{Comparison of Bias Estimates}\label{Tab_Gaussian_model}
\begin{center}
\begin{tabular}{c|cccc|cccc}
         & \multicolumn{4}{|c|}{Data: Gauss}  & \multicolumn{4}{|c}{Data: Laplace} \\
         &   25  &  100  &  400  & 1,600 &   25  &  100  &  400  & 1,600 \\
\hline
 True       & 2.272 & 2.061 & 2.015 & 2.006 & 3.874 & 3.574 & 3.516 & 3.502 \\
 AIC        & 2.000 & 2.000 & 2.000 & 2.000 & 2.000 & 2.000 & 2.000 & 2.000 \\
 $IJ^{-1}$  & 2.000 & 2.000 & 2.000 & 2.000 & 3.500 & 3.500 & 3.500 & 3.500 \\
 $\hat{IJ}^{-1}$ & 1.884 & 1.970 & 1.993 & 1.998 & 2.594 & 3.166 & 3.402 & 3.474 \\
 $C_n$      & 2.273 & 2.062 & 2.015 & 2.004 & 4.040 & 3.635 & 3.534 & 3.508\\
 $B_n$     & 2.226 & 2.037 & 2.008 & 2.002 & 3.433 & 3.331 & 3.431 & 3.479\\
\hline
\end{tabular}
\end{center}
\end{table}

Relatively good values are obtained by most of the methods when the true distribution is Gaussian. 
However, AIC and $IJ^{-1}$ underestimate compared to ^^ ^^ True" when the number of data $N$ is small. It should be noted that $IJ^{-1}$, which estimates higher-order moments from the data, has a larger discrepancy and underestimates than AIC. 
Also, in this case, $C_n$ is a small sample adjustment of AIC, but it approximates ^^ ^^ True" value very well. The bootstrap estimate $B_n$ is also a relatively good estimate.

When the true distribution is Laplace, there is a large difference between the True and AIC values, indicating that the Gauss distribution model does not provide a good approximation; $IJ^{-1}$ gives a reasonable estimate. However, the estimates $\hat{IJ}^{-1}$ is a significant underestimate for small numbers of data. This is due to the fact that the fourth-order moments $\mu_4$ estimated from the data are quite small, 5.31 and 5.82 for N=25 and 100, respectively. The variances of these moments are 75.33 and 22.99,respectively, which are very large, indicating that the estimation is unstable.
$C_n$ gives a good approximation when $N$ is large, but is considerably overestimated when $N$ is small. In contrast, the bootstrap estimate is conversely underestimated in the small sample situation.

\section{Laplace Distribution Model}

In this section, we consider the Laplace distribution model defined by
\begin{eqnarray}
   f(x|\xi ,\tau ) &=& \frac{1}{2\tau } \exp \left( - \frac{|x-\xi |}{\tau} \right),
\end{eqnarray}
where $\xi$ is the location parameter and $\tau$ is the scale parameter.
Given $N$ independent observations $x_1,\ldots ,x_N$, the log-likelihood of the Laplace distribution model is defined by
\begin{eqnarray}
  \ell (\xi, \tau ) &=& -N\log 2\tau - \frac{1}{\tau} \sum_{n=1}^N |x_n-\xi |.
\end{eqnarray}
Then the maximum likelihood estimates of the Laplace distribution model are given by the sample median and the mean absolute deviation from the median, respectively,
\begin{eqnarray}
   \hat{\xi}  = \textrm{median} ( x_1,\ldots ,x_N ), \qquad
   \hat{\tau} = \frac{1}{N} \sum_{n=1}^N \bigl|x_n -\hat{\xi}\bigr|.
\end{eqnarray}
The maximum log-likelihood is given by
\begin{eqnarray}
  \ell (\hat{\xi}, \hat{\tau} ) &=& -N\log 2\hat{\tau} - N.
\end{eqnarray}

For the Laplace distribution model, the Fisher information matrix and the expected Hessian are
given by equations (\ref{Eq_Fisher-Information_Laplace}) and (\ref{Eq_Hessian_Laplace}), respectively as
\begin{eqnarray}
I(\theta_0) = \left[ \begin{array}{cc} 
            \frac{1}{\tau^2} & 0  \\
            0 & \frac{1}{\tau^2} - 2\frac{\delta_1}{\tau^3} + \frac{\mu_2}{\tau^4} \end{array} \right] ,
\quad
J(\theta_0) = \left[ \begin{array}{cc} 0 & 0  \\
                 0 & -\frac{1}{\tau^2} + \frac{2\delta_1}{\tau^3} \end{array} \right] .
\end{eqnarray}
The bias correction term tr$\{I(\theta_0)J(\theta_0)^{-1}\}$ in the TIC cannot be calculated because the expected Hessian is a singular matrix.

\subsection{For Laplace Distribution Data}
Here we consider the case that the true model generating the data is the
Laplace distribution model with location parameter $\mu$ and the scale parameter $\sigma$,
\begin{eqnarray}
   g(x|\mu ,\sigma ) &=& \frac{1}{2\sigma } \exp \left( - \frac{|x-\mu |}{\sigma} \right).
\end{eqnarray}
The logarithm of the Laplace distribution model and its expected values with respect to the true distribution, called the expected log-likelihood, are defined by
\begin{eqnarray}
  \log f(x|\xi,\tau) &=& -\log 2\tau - \frac{|x-\xi |}{\tau}  \nonumber\\
   E_L[\log f(x|\hat{\xi} ,\hat{\tau} )] 
   &=& -\log 2\hat{\tau} - \frac{ E_L\bigl[\, |x-\hat{\xi} |\,\bigr] }{\hat{\tau}} \nonumber \\
   &=& -\log 2\hat{\tau} - \frac{\sigma}{\hat{\tau}}\exp\left\{-\frac{|\hat{\xi}-\mu|}{\sigma}\right\}
       - \frac{|\hat{\xi}-\mu|}{\hat{\tau}}, \nonumber 
\end{eqnarray}
where $E_L$ denotes the expectation with respect to the true distribution $g(x|\mu,\sigma)$.
The true parameters $\mu$ and $\sigma$ maximize the expected log-likelihood and the mximum value is given by
\begin{eqnarray}
   E_L[\log f(x|\mu ,\sigma )] &=& -\log 2\sigma - 1.  \nonumber   
\end{eqnarray}

The asymptotic distribution of the maximum likelihood estimates of the Laplace
distribution model is given by
\begin{eqnarray}
  \hat{\xi} \sim N\left(\mu,\frac{\tau^2}{N}\right),\qquad
  \hat{\sigma}\sim N\left(\left(1-\frac{1}{2N}\right)\tau,\frac{\tau^2}{N}\right).  
\end{eqnarray}
Using these distributions and the asymptotic independence of $\hat{\mu}$ and $\hat{\tau}$,
and equations (\ref{Eq_Laplace-Laplace_E[tau/sigma]}), (\ref{Eq_Laplace-Laplace_E[exp(-|X|]}) and (\ref{Eq_Laplace-Laplace_E[|xi-mu|]}),  the bias correction term is evaluated as
\begin{eqnarray}
C_N &=& E \biggl[ \ell(\hat{\xi},\hat{\sigma})
                      -  N E_{L}  \log f(Y|\hat{\xi},\hat{\sigma}) \biggr] \nonumber \\
  &=& N E \left[ \frac{\sigma}{\hat{\tau}}\left\{ \exp\left(-\frac{|\hat{\xi}-\mu|}{\sigma}\right)
       - \frac{|\hat{\xi}-\mu|}{\sigma} \right\} - 1 \right]  \nonumber \\
  &\approx& N E \left[ \frac{\sigma}{\hat{\tau}}\right] 
        \left\{ E\left[\exp\left(-\frac{|\hat{\xi}-\mu|}{\sigma}\right)\right] 
      - \frac{E[\,|\hat{\xi}-\mu|\,]}{\sigma}\right\}  - N   \nonumber \\
  &\approx&  N \left( 1 + \frac{3}{2N} + \frac{19}{4N^2} + \cdots \right)
        \left\{ \exp\left( \frac{1}{2N} \right)\textrm{erfc}\left( \frac{1}{\sqrt{2N}} \right) 
           - \frac{\sqrt{2}}{\sqrt{N}} \right\} - N \nonumber \\
  &\approx& N \left( 1 + \frac{3}{2N} + \frac{19}{4N^2} + \cdots \right)
        \biggl( 1 + \frac{1}{2N} - \frac{\sqrt{2}}{3N\sqrt{N\pi}} + \frac{1}{8N^2}
                  + \cdots \biggr) - N \nonumber\\
  &\approx& 2 - \frac{\sqrt{2}}{3\sqrt{\pi N}} + \frac{45}{8N} .
\end{eqnarray}

\subsection{For Gaussian Distribution Data}
In this subsection, we assume that the true model generating the data is the
Gaussian distribution with mean $\mu$ and the variance $\sigma^2$,
\begin{eqnarray}
   g(x|\mu ,\sigma^2 ) &=& \frac{1}{\sqrt{2\pi\sigma^2} } \exp \left\{ - \frac{(x-\mu)^2}{2\sigma^2} \right\}.
\end{eqnarray}
Using (\ref{Eq_E[|X|]_Gauss}), the logarithm of the Laplace distribution model and the expected values with respect to the true distribution are defined by
\begin{eqnarray}
  \log f(x|\mu,\tau) &=& -\log 2\tau - \frac{|x-\xi |}{\tau}  \nonumber\\
   E_G[\log f(x|\xi ,\tau )] &=& -\log 2\tau - \frac{E_G\left[\, |x-\xi |\,\right] }{\tau} \nonumber \\
   &=& -\log 2\tau - \frac{(\xi-\mu)}{\tau}\textrm{erf}\left( \frac{\xi-\mu}{\sqrt{2\sigma^2}} \right) 
       - \frac{1}{\tau}\sqrt{\frac{2\sigma^2}{\pi}}\exp\left\{-\frac{(\xi-\mu)^2}{2\sigma^2} \right\}, \nonumber 
\end{eqnarray}
where erfc$(x)$ is the complementary error function defined by
erfc$(x) = 1 - \textrm{erf}(x) =\frac{2}{\sqrt{\pi}} \int_x^{\infty} e^{-y^2} dy$.
$\xi$ and $\tau$ that maximize the expected log-likelihood is given by
$\xi^{\ast} = \mu$ and $\tau^{\ast} = \sqrt{\frac{2}{\pi}}\sigma$ and the the maximum of the 
expected log-likelihood is obtained as
\begin{eqnarray}
   E_G[\log g(x|\xi_0 ,\tau_0 )] &=& -\log 2\tau_0  - 1.  \nonumber 
\end{eqnarray}

Using the asymptotic distribution of the maximum likelihood estimates
\begin{eqnarray}
 \left(\begin{array}{c} \hat{\xi} \\ \hat{\tau} \end{array} \right) 
&=& N\left( \left[ \begin{array}{c} \mu \\ \sqrt{\frac{2}{\pi}}\sigma  \end{array} \right], 
        \left[ \begin{array}{cc} \frac{\pi}{2n}\sigma^2  & 0 \\
               0 & \frac{\pi-2}{\pi n}\sigma^2  \end{array}\right] \right),
\end{eqnarray}
and equations (\ref{Eq_Gauss-Laplace_E[tau-sigma]}), (\ref{Eq_Gauss-Laplace_E[exp(-|X|)]}), (\ref{Eq_Gauss-Laplace_E[|xi-mu|]}) and (\ref{Eq_Gauss-Laplace_sqrt(2n/2n+pi)}) in Appendix D, we obtain
%
\begin{small}\begin{eqnarray}
\lefteqn{ E_{\hat{G}}\left[ \ell (\hat{\xi},\hat{\sigma} ) - E_L[\log f(x|\hat{\xi},\hat{\sigma} )] \right] }
\nonumber \\
 &=& E_{\hat{G}}\left[ \frac{1}{\hat{\sigma}}\sqrt{\frac{2\sigma^2}{\pi}}\exp\left\{-\frac{(\hat{\xi}-\mu)^2}{2\sigma^2}\right\}
      +  \frac{\hat{\xi}-\mu}{\hat{\sigma}}\textrm{erf}\left( \frac{\hat{\xi}-\mu}{\sqrt{2\sigma^2}} \right)  - 1 \right]  \nonumber \\
 &\approx& E_{\hat{G}}\left[ \frac{\tau}{\hat{\sigma}}\right] 
    \left( \frac{2}{\sqrt{\pi}}E_{\hat{G}}\left[ \exp\left\{-\frac{(\hat{\xi}-\mu)^2}{\tau^2}\right\} \right]
  + 2 E_{\hat{G}}\left[ \frac{\hat{\xi}-\mu}{\sqrt{2}\sigma}\textrm{erf}\left( \frac{\hat{\xi}-\mu}{\sqrt{2}\sigma} \right) \right] \right) 
            - 1   \nonumber \\
 &\approx& \frac{\sqrt{\pi}}{2} \left( 1 + \frac{(\pi-2)}{2n} + \frac{3(\pi-2)^2}{4n^2} + \frac{15(\pi-2)^3}{8n^3} \right)
           \left\{ \frac{2}{\sqrt{\pi}}\frac{\sqrt{2n}}{\sqrt{2n+\pi}}
      + \frac{\sqrt{\pi}}{\sqrt{n(2n+\pi)}}  \right\} - 1 \nonumber \\
 &\approx&  \left\{ 1 + \frac{(\pi-2)}{2n} + \frac{3(\pi-2)^2}{4n^2} + \frac{15(\pi-2)^3}{8n^3} \right\}
   \!\biggl\{1 + \frac{(2\sqrt{2}-1)\pi}{4n} - \frac{4\sqrt{2}-3\pi^2}{32n^2} + \frac{6\sqrt{2}-5}{4}\pi + 3 \biggr\} - 1   \nonumber \\
 &\approx&   \biggl(\frac{2\sqrt{2}+1}{4}\pi - 1 \biggr)n^{-1} 
   + \biggl( \frac{4\sqrt{2}+31}{32}\pi^2  - \frac{2\sqrt{2}+13}{4}\pi + 3 \biggr) n^{-2}.
\end{eqnarray}\end{small}

\subsection{Numerical Example}
Table \ref{Tab_Laplace_model} shows the bias correction term evaluated by various methods.
^^ ^^ True", AIC, $IJ^{-1}$, $C_n$ and $B_n$ are defined as in Table \ref{Tab_Gaussian_model}.
Unlike in the Gaussian model, $IJ^{-1}$ cannot be defined for Laplace distribution model.
^^ ^^ True" value of the bias correction term is similar whether the true model is the Laplace model or the Gauss model. This may be due to the fact that, unlike the case in the previous section, the Laplace distribution model provides a relatively good approximation even when the true distribution is Gaussian.

\begin{table}[h]
\caption{Comparison of Bias Estimates for Laplace Distribution Model}\label{Tab_Laplace_model}
\begin{center}
\begin{tabular}{c|cccc|cccc}
         & \multicolumn{4}{|c|}{Data: Gauss}  & \multicolumn{4}{|c}{Data: Laplace} \\
         &   25  &  100  &  400  & 1,600 &   25  &  100  &  400  & 1,600 \\
\hline
 True      & 2.254 & 2.151 & 2.149 & 2.120 & 2.274 & 2.096 & 2.043 & 2.021 \\
 AIC       & 2.000 & 2.000 & 2.000 & 2.000 & 2.000 & 2.000 & 2.000 & 2.000 \\
 $IJ^{-1}$ &  ---  &  ---  &  ---  &  ---  &  ---  &  ---  &  ---  &  ---  \\
 $C_n$     & 2.108 & 2.032 & 2.012 & 2.008 & 2.172 & 2.030 & 2.001 & 1.997 \\
 $B_n$     & 2.333 & 2.176 & 2.150 & 2.142 & 2.158 & 2.075 & 2.061 & 2.027 \\
\hline
\end{tabular}
\end{center}
\end{table}

In both cases, the AIC is a relatively good approximation, and good approximations are obtained for the other bias correction terms. 
When the true model is a Laplace distribution, i.e., the model includes the true model, both $C_n$ and $B_n$ give good approximations of the True bias, but both are slightly underestimated.
On the other hand, when the true model is a Gauss distribution, $C_n$ is slightly underestimated and $B_n$ is slightly overestimated.

\section{Concuding Remarks}
The results of this paper can be summarized as follows:
\begin{itemize}
\item AIC underestimates the true bias, but has the advantage that it is independent of the data and thus provides a stable value without any additional calculations.
\item  The $IJ^{-1}$ of the TIC gives a good bias correction value for Gaussian models, even if the true model is Laplace distributed. It should be noted, however, that estimating $IJ^{-1}$ from the data does not necessarily improve the AIC in real situations where the higher-order moments are unknown. It also cannot be defined in the case of the Laplace distribution model because the expected value of Hessian matrix is singular.
Translated with DeepL.com (free version)
\item $C_N$, which evaluates the bias directly, automatically makes a finite correction and gives a relatively good estimate. However, it requires a series approximation and is likely to be difficult to apply in the case of complex models.
\item The bootstrap bias correction term $B_N$ is based on the bootstrap method and is computationally expensive, but has the advantage that it is easy to implement and can obtain a relatively good bias correction for any model or distribution.
\end{itemize}

\newpage
\begin{center}\begin{Large}
\textbf{Appendix}
\end{Large}\end{center}

\appendix
\section{Fisher Information Matrices and Expected Hessians for the Gaussian and Laplace Distribution Models}
\subsection{Gaussian Distribution Model}

For the Gaussian distribution model, 
\begin{eqnarray}
f(x|\xi,\tau^2) = \frac{1}{\sqrt{2\pi\tau^2}} \exp\left\{-\frac{(x-\xi)^2}{2\tau^2} \right\},
\end{eqnarray}
where $\xi$ is the mean and $\tau^2$ is the variance,
the logrithm of the density function and its derivatives are given by
\begin{eqnarray}
  \log f(x|\xi ,\tau^2 ) &=& -\frac{1}{2}\log 2\pi\tau^2 - \frac{(x-\xi )^2}{2\tau^2} \nonumber \\
  \frac{\partial}{\partial\xi}\log f(x|\xi ,\tau^2 ) &=& \frac{x-\xi}{\tau^2} \nonumber \\
  \frac{\partial}{\partial\tau^2}\log f(x|\xi ,\tau^2 ) 
        &=& -\frac{1}{2\tau^2} + \frac{(x-\xi)^2}{2\tau^4} \nonumber \\
  \frac{\partial^2}{\partial\xi\partial\xi}\log f(x|\xi ,\tau^2 ) &=& -\frac{1}{\tau^2} \nonumber \\  \frac{\partial^2}{\partial\xi\partial\tau^2}\log f(x|\xi ,\tau^2 ) &=& -\frac{x-\xi}{\tau^4} \nonumber \\
  \frac{\partial^2}{\partial\tau^2\partial\tau^2}\log f(x|\xi ,\tau^2 ) 
        &=& \frac{1}{2\tau^4} - \frac{(x-\xi)^2}{\tau^6}. \nonumber 
\end{eqnarray}


Taking expectation with respect to the true distribution generating the data, we obtain
\begin{eqnarray}
E\left[ \frac{\partial^2}{\partial\xi\partial\xi}\log f(x|\xi ,\tau^2 )\right]_{\xi=\mu}  
     &=& -\frac{1}{\tau^2} \nonumber \\
E\left[  \frac{\partial^2}{\partial\xi\partial\sigma^2}\log f(x|\xi ,\tau^2 ) \right]_{\xi =\mu} 
     &=& - \frac{E\left[x-\xi\right]}{\tau^4} = 0 \nonumber \\
E\left[  \frac{\partial^2}{\partial\sigma^2\partial\sigma^2}\log f(x|\xi ,\tau^2 ) \right]_{\xi =\mu} 
     &=& \frac{1}{2\tau^4} - \frac{E[(x-\xi)^2]}{\tau^6}
       =-\frac{1}{2\tau^4}  \nonumber \\
E\left[ \frac{\partial}{\partial\xi}\log f(x)\frac{\partial}{\partial\xi}\log f(x) \right]_{\xi =\mu} 
   &=& \frac{1}{\tau^4} E\bigl[ (x-\xi)^2 \bigr] 
    =  \frac{1}{\tau^2} \nonumber \\
E\left[ \frac{\partial}{\partial\xi}\log f(x) \frac{\partial}{\partial\tau^2}\log f(x) \right]_{\xi =\mu} 
     &=& -\frac{E_G\left[ (x-\xi) \right]}{2\tau^4}  
         +\frac{E_G\left[ (x-\xi)^3 \right]}{2\tau^6}  
     = \frac{\mu_3}{2\tau^6} \nonumber \\
E\left[ \frac{\partial}{\partial\tau^2}\log f(x) \frac{\partial}{\partial\tau^2}\log f(x) \right]_{\xi =\mu} 
     &=& \frac{1}{4\tau^4} - \frac{E[(x-\xi)^2}{2\tau^6} + \frac{E[(x-\xi)^4]}{4\tau^8}
       =-\frac{1}{4\tau^4} + \frac{\mu_4}{4\tau^8},  \nonumber 
\end{eqnarray}
where $\mu_k$ is the central moment of order $k$, i.e, $\mu_k = E[\,(x-\xi)^k \,]$.
Using these, the Fisher information, $I(\theta_0)$ and the negative of the Hessian matrix, $J(\theta_0)$ are given by
\begin{eqnarray}
I(\theta_0) &=& \left[ \begin{array}{cc} 
            \frac{1}{\tau^2} & \frac{\mu_3}{2\tau^6}  \label{Eq_Fisher-Information_Gauss}\\
            \frac{\mu_3}{2\sigma^6} & \frac{\mu_4}{4\sigma^8} - \frac{1}{4\sigma^4} \end{array} \right] \\
J(\theta_0) &=& \left[ \begin{array}{cc} \frac{1}{\tau^2} & 0  \\
                                         0 & \frac{1}{2\tau^4} \end{array} \right] . \label{Eq_Hessian_Gauss}
\end{eqnarray}

\subsection{Laplace Distribution Model}

For the Laplace distribution model, 
\begin{eqnarray}
  f(x|\xi,\tau) = \frac{1}{2\tau} \exp\left( -\frac{|x-\xi|}{\tau}\right),
\end{eqnarray}
where $\mu$ is the central parameter and $\tau$ is the scale parameter,
the logrithm of the density function and its derivatives are given by
\begin{eqnarray}
   f(x|\xi,\tau) &=& \frac{1}{2\tau} \exp\left( -\frac{|x-\xi|}{\tau} \right) \nonumber \\
 \log f(x|\xi,\tau) &=& -\log 2\tau -\frac{|x-\xi|}{\tau}  \nonumber \\
  \frac{\partial}{\partial\tau}\log f(x|\xi,\tau) &=& -\frac{1}{\tau} + \frac{|x-\xi|}{\tau^2}  \nonumber \\
  \frac{\partial}{\partial\xi}\log f(x|\xi,\tau) &=& \frac{x-\xi}{\tau|x-\xi|} = 
      \left\{ \begin{array}{cl}  \frac{1}{\tau} & x>\xi \\
                                -\frac{1}{\tau} & x<\xi \end{array} \right. \nonumber \\
  \frac{\partial^2}{\partial\xi\partial\xi}\log f(x|\xi,\tau) &=& 0  \nonumber \\
  \frac{\partial^2}{\partial\tau\partial\tau}\log f(x|\xi,\tau) 
      &=& \frac{1}{\tau^2} - \frac{2|x-\xi|}{\tau^3}  \nonumber \\
  \frac{\partial^2}{\partial\xi\partial\tau}\log f(x|\xi,\tau) &=& 
      \left\{ \begin{array}{cl} -\frac{1}{\tau^2} & x>\xi \\
                                 \frac{1}{\tau^2} & x<\xi \end{array} \right.  . \nonumber 
\end{eqnarray}

Taking expectation with respect to the true distribution generating data using
$g(x-\xi)=g(\xi-x)$, $y=x-\xi$, $\mu_k = E[(x-\xi)^k]$, $\delta_k =E[\,|x-\xi|^k\,]$,
we obtain
\begin{small}
\begin{eqnarray}
E\left[ \frac{\partial^2}{\partial\xi\partial\xi}\log f(x|\xi ,\tau )\right]_{\xi =\mu}  
   \!\!  &=& 0  \nonumber \\
E\left[  \frac{\partial^2}{\partial\xi\partial\tau}\log f(x|\xi ,\tau ) \right]_{\xi =\mu} 
   \!\!  &=& \!\int_{0}^{\infty} \frac{1}{\tau} g(y) dy
       - \!\int_{-\infty}^{0} \!\frac{1}{\tau^2} g(y) dy 
      = \lim_{t\rightarrow -\infty} \int_0^t \left(\frac{1}{\tau^2}-\frac{1}{\tau^2}\right)\!dy = 0
 \nonumber \\  
E\left[  \frac{\partial^2}{\partial\tau\partial\tau}\log f(x|\xi ,\tau ) \right]_{\xi =\mu} 
   \!\!  &=& \frac{1}{\tau^2} - \frac{2E[\,|x-\xi|\,]}{\tau^3}
     =   \frac{1}{\tau^2} - \frac{2\delta_1}{\tau^3} \nonumber \\
E\left[ \frac{\partial}{\partial\xi}\log f(x)\frac{\partial}{\partial\xi}\log f(x) \right]_{\xi =\mu} 
   &=& \frac{1}{\tau^2} \nonumber \\
E\left[ \frac{\partial}{\partial\xi}\log f(x) \frac{\partial}{\partial\tau}\log f(x) \right]_{\xi =\mu} 
  \!\!   &=& \!\!-\!\int_{\xi}^{\infty} \!\!\left(\frac{1}{\tau^2} - \!\frac{|x-\xi|}{\tau^3} \right)\! g(x) dx
       + \!\int_{-\infty}^{\xi} \!\!\left(\frac{1}{\tau^2} - \frac{|x-\xi|}{\tau^3} \right)\! g(x) dx = 0
 \nonumber \\
E\left[ \frac{\partial}{\partial\tau}\log f(x) \frac{\partial}{\partial\tau}\log f(x) \right]_{\xi =\mu} 
  \!\!   &=& \frac{1}{\tau^2} - \frac{2E[\,|x-\xi|\,]}{\tau^3} + \frac{E[\,(x-\xi)^2\,]}{\tau^4}
       = \frac{1}{\tau^2} - \frac{2\delta_1}{\tau^3} + \frac{\mu_2}{\tau^4}  \nonumber 
\end{eqnarray}\end{small}

Therefore, the Fisher information matrix and the negative of the expected Hessian are respectively given by
\begin{eqnarray}
I(\theta_0) &=& \left[ \begin{array}{cc} 
            \frac{1}{\tau^2} & 0  \\
            0 & \frac{1}{\tau^2} - \frac{2\delta_1}{\tau^3} + \frac{\mu_2}{\tau^4} \end{array} \right] 
\label{Eq_Fisher-Information_Laplace} \\
J(\theta_0) &=& \left[ \begin{array}{cc} 0 & 0  \\
                 0 & -\frac{1}{\tau^2} + \frac{2\delta_1}{\tau^3} \end{array} \right] .
\label{Eq_Hessian_Laplace} 
\end{eqnarray}

\section{Expectation with respect to the Gaussian distribution}

\subsection{Expectation of $(x-\xi)^k$}
\begin{small}
\begin{eqnarray}
 E_G\left[ (x-\xi) \right] &=& \frac{1}{\sqrt{2\pi\sigma^2}} \int_{-\infty}^{\infty} (x-\xi) \exp\left\{ -\frac{(x-\mu)^2}{2\sigma^2} \right\}dx  \nonumber  \\
&=& \frac{1}{\sqrt{\pi}} \int_{-\infty}^{\infty} (\mu-\xi+\sqrt{2\sigma^2} y) \exp\left( -y^2 \right)dy \nonumber \\
&=& \mu-\xi  \nonumber  \\
 E_G\left[\, (x-\xi )^2\,\right]  &=& \frac{1}{\sqrt{2\pi\sigma^2}} \int_{-\infty}^{\infty} (x-\xi)^2 \exp\left\{ -\frac{(x-\mu)^2}{2\sigma^2} \right\}dx  \nonumber  \\
&=& \frac{1}{\sqrt{\pi}} \int_{-\infty}^{\infty} (\mu-\xi+\sqrt{2\sigma^2} y)^2 \exp\left( -y^2 \right)dy \nonumber \\
&=& \frac{(\mu-\xi)^2}{\sqrt{\pi}} \!\int_{-\infty}^{\infty} \!\!\!\! \exp\left( -y^2 \right)dy 
  + \frac{2\sqrt{2\sigma^2}(\mu-\xi)}{\sqrt{\pi}} \!\int_{-\infty}^{\infty} \!\!\!y \exp\left( -y^2 \right)dy 
  + \frac{2\sigma^2}{\sqrt{\pi}} \!\int_{-\infty}^{\infty} \!\!\!y^2 \exp\left( -y^2 \right)dy  \nonumber \\
&=& (\mu-\xi)^2 + \sigma^2 \nonumber\\
 E_G\left[ (x-\xi)^3\right] &=&  \frac{1}{\sqrt{2\pi\sigma^2}} \int_{-\infty}^{\infty} (x-\xi)^3 \exp\left\{ -\frac{(x-\mu)^2}{2\sigma^2} \right\}dx  \nonumber  \\
&=& \frac{1}{\sqrt{\pi}} \!\int_{-\infty}^{\infty} (\mu-\xi+\sqrt{2\sigma^2} y)^3 \exp\left( -y^2 \right)dy \nonumber \\
&=& \frac{(\mu-\xi)^3}{\sqrt{\pi}} \!\int_{-\infty}^{\infty} \!\!\exp\left( -y^2 \right)dy 
  + \frac{3\sqrt{2\sigma^2}(\mu-\xi)^2}{\sqrt{\pi}} \!\int_{-\infty}^{\infty} \!\!y \exp\left( -y^2 \right)dy  \nonumber \\
&&+ \frac{6\sigma^2(\mu-\xi)}{\sqrt{\pi}} \!\int_{-\infty}^{\infty} \!\!y^2 \exp\left( -y^2 \right)dy 
  + \frac{2\sqrt{2}\sigma^3}{\sqrt{\pi}} \!\int_{-\infty}^{\infty} \!\!y^3 \exp\left( -y^2 \right)dy  \nonumber \\
&=& (\mu-\xi)^3 + 3\sigma^2(\mu-\xi)  \nonumber  \\
 E_G\left[ (x-\xi)^4\right] &=& \frac{1}{\sqrt{2\pi\sigma^2}} \int_{-\infty}^{\infty} (x-\xi)^4 \exp\left\{ -\frac{(x-\mu)^2}{2\sigma^2} \right\} dx  \nonumber  \\
&=& \frac{1}{\sqrt{\pi}} \!\int_{-\infty}^{\infty} (\mu-\xi+\sigma y)^4 \exp\left( -y^2 \right)dy \nonumber \\
&=& (\mu-\xi)^4 + 6\sigma^2(\mu-\xi)^2 + 3\sigma^4  \nonumber 
\end{eqnarray}\end{small}
In summary
\begin{eqnarray}
E_G\Bigl[\, x-\xi\, \Bigr] &=& \mu-\xi \nonumber \\
E_G\left[(x-\xi)^2 \right] &=& (\mu-\xi)^2 + \sigma^2  \nonumber \\
E_G\left[(x-\xi)^3 \right] &=& (\mu-\xi)^3 + 3(\mu-\xi)\sigma^2  \nonumber \\
E_G\left[(x-\xi)^4 \right] &=& (\mu-\xi)^4 + 6\sigma^2(\mu-\xi)^2 + 3\sigma^4 \nonumber
\end{eqnarray}

\subsection{Expectation of $|x-\xi|^k$}
For simplicity, we denote $a=\frac{\xi-\mu}{\sqrt{2\sigma^2}}$.
\begin{eqnarray}
 E_G\left[\, |x-\xi|\, \right] &=& \frac{1}{\sqrt{2\pi\sigma^2}} \int_{-\infty}^{\infty} |x-\xi| \exp\left\{ -\frac{(x-\mu)^2}{2\sigma^2} \right\}dx  \nonumber  \\
 &=& \frac{1}{\sqrt{2\pi\sigma^2}} \int_{-\infty}^{\xi} (\xi -x) \exp\left\{ -\frac{(x-\mu)^2}{2\sigma^2} \right\}dx  
  +  \frac{1}{\sqrt{2\pi\sigma^2}} \int_{\xi}^{\infty} (x-\xi) \exp\left\{ -\frac{(x-\mu)^2}{2\sigma^2} \right\}dx 
\nonumber  \\
 &=& \frac{1}{\sqrt{\pi}} \int_{-\infty}^{a} (\xi -\mu-\sqrt{2\sigma^2}y) \exp\left( -y^2 \right)dy  
  +  \frac{1}{\sqrt{\pi}} \int_{a}^{\infty} (\mu-\xi+\sqrt{2\sigma^2}y) \exp\left( -y^2 \right)dy 
\nonumber  \\
&=& \frac{1}{2}(\xi-\mu)(\textrm{erf}(a)+1) + \frac{1}{2}\sqrt{\frac{2\sigma^2}{\pi}} e^{-a^2}
   + \frac{1}{2}(\mu-\xi)\textrm{erfc}(a) + \frac{1}{2}\sqrt{\frac{2\sigma^2}{\pi}} e^{-a^2} 
   \nonumber \\
&=& (\xi-\mu)\textrm{erf}\left( \frac{\xi-\mu}{\sqrt{2\sigma^2}}\right) 
    + \sqrt{\frac{2\sigma^2}{\pi}} \exp\left\{-\frac{(\xi-\mu)^2}{2\sigma^2} \right\} \nonumber\\
&=& (\xi-\mu)\textrm{erf}\left( \frac{\xi-\mu}{2\tau}\right) 
    + \frac{2\tau}{\sqrt{\pi}} \exp\left\{-\frac{(\xi-\mu)^2}{4\tau^2} \right\}   \label{Eq_E[|X|]_Gauss} \\
\end{eqnarray}
%

\section{Expectation with respect to the Laplace distribution}

\subsection{Expectation of $(x-\xi)^k$}
\begin{eqnarray}
 E_L\left[ (x-\xi) \right] &=& \frac{1}{2\sigma} \int_{-\infty}^{\infty} (x-\xi) \exp\left( -\frac{|x-\mu|}{\sigma} \right)dx  \nonumber  \\
&=& \frac{1}{2} \!\int_{-\infty}^{\infty} (\mu-\xi+\sigma y) \exp\left( -|y| \right)dy \nonumber \\
&=& \frac{\mu-\xi}{2} \int_{-\infty}^{\infty} \exp\left( -|y| \right)dy 
  + \frac{\sigma}{2} \int_{-\infty}^{\infty} y \exp\left( -|y| \right)dy   \nonumber\\
&=& \mu-\xi    \\
 E_L\left[\, (x-\xi )^2\,\right]  &=& \frac{1}{2\sigma} \int_{-\infty}^{\infty} (x-\xi)^2 \exp\left( -\frac{|x-\mu|}{\sigma} \right)dx  \nonumber  \\
&=& \frac{1}{2} \!\int_{-\infty}^{\infty} (\mu-\xi+\sigma y)^2 \exp\left( -|y| \right)dy \nonumber \\
&=& \frac{1}{2}(\mu-\xi)^2 \!\int_{-\infty}^{\infty} \!\!\!\! \exp\left( -|y| \right)dy 
  + \sigma(\mu-\xi) \!\int_{-\infty}^{\infty} \!\!\!y \exp\left( -|y| \right)dy 
  + \frac{\sigma^2}{2} \!\int_{-\infty}^{\infty} \!\!\!y^2 \exp\left( -|y| \right)dy  \label{Eq_X2_Laplace} \nonumber\\
&=& (\mu-\xi)^2 + 2\sigma^2   \label{Eq_expectation_of_x^2_Laplace} \\
 E_L\left[ (x-\xi)^3\right] &=&  \frac{1}{2\sigma} \int_{-\infty}^{\infty} (x-\xi)^3 \exp\left( -\frac{|x-\mu|}{\sigma} \right)dx  \nonumber  \\
&=& \frac{1}{2} \!\int_{-\infty}^{\infty} (\mu-\xi+\sigma y)^3 \exp\left( -|y| \right)dy \nonumber \\
&=& \frac{(\mu-\xi)^3}{2} \!\int_{-\infty}^{\infty} \!\!\exp\left( -|y| \right)dy 
  + \frac{3\sigma(\mu-\xi)^2}{2} \!\int_{-\infty}^{\infty} \!\!y \exp\left( -|y| \right)dy  \nonumber \\
&&+ \frac{3\sigma^2(\mu-\xi)}{2} \!\int_{-\infty}^{\infty} \!\!y^2 \exp\left( -|y| \right)dy 
  + \frac{\sigma^3}{2} \!\int_{-\infty}^{\infty} \!\!y^3 \exp\left( -|y| \right)dy   \\
&=& (\mu-\xi)^3 + 6\sigma^2(\mu-\xi)    \\
 E_L\left[ (x-\xi)^4\right] &=& \frac{1}{2\sigma} \int_{-\infty}^{\infty} (x-\xi)^4 \exp\left( -\frac{|x-\mu|}{\sigma} \right)dx  \nonumber  \\
&=& \frac{1}{2} \!\int_{-\infty}^{\infty} (\mu-\xi+\sigma y)^4 \exp\left( -|y| \right)dy \nonumber \\
&=& (\mu-\xi)^4 + 12\sigma^2(\mu-\xi)^2 + 24\sigma^4   
\end{eqnarray}
In summary, we have
\begin{eqnarray}
E_L\Bigl[ x-\xi \Bigr] &=& \mu-\xi \nonumber \\
E_L\left[(x-\xi)^2 \right] &=& (\mu-\xi)^2 + 2\sigma^2  \nonumber \\
E_L\left[(x-\xi)^3 \right] &=& (\mu-\xi)^3 + 6(\mu-\xi)\sigma^2  \nonumber \\
E_L\left[(x-\xi)^4 \right] &=& (\mu-\xi)^4 + 12\sigma^2(\mu-\xi)^2 + 24\sigma^4 . \nonumber
\end{eqnarray}

\subsection{Expectation of $|x-\xi|^k$}
Assuming  that $\mu > \xi$ and $a=\frac{\xi-\mu}{\sigma}$,
\begin{eqnarray}
 E_L\Bigl[\, |x-\xi|\, \Bigr] 
  &=& \frac{1}{2\sigma} \int_{-\infty}^{\xi} (\xi-x) \exp\left( -\frac{|x-\mu|}{\sigma} \right)dx
   + \frac{1}{2\sigma} \int^{\infty}_{\xi}( x-\xi) \exp\left( -\frac{|x-\mu|}{\sigma} \right)dx \nonumber\\
  &=& \frac{1}{2\sigma} \int_{-\infty}^{\xi} (\xi-x) \exp\left( -\frac{\mu-x}{\sigma} \right)dx
   + \frac{1}{2\sigma} \int_{\xi}^{\mu}( x-\xi) \exp\left( -\frac{\mu-x}{\sigma} \right)dx \nonumber\\
  & & + \frac{1}{2\sigma} \int_{\mu}^{\infty}( x-\xi) \exp\left( -\frac{x-\mu}{\sigma} \right)dx \nonumber\\
&=& \frac{1}{2}\left\{ (\mu-\xi) + \sigma \right\}
  + \frac{1}{2}\left\{ (\xi-\mu)\exp\left( \frac{\xi-\mu}{\sigma} \right) 
           - \sigma\left( \frac{\xi-\mu}{\sigma} - 1 \right) \exp\left(\frac{\xi-\mu}{\sigma}\right)\right\} \nonumber\\
&& + \frac{1}{2}\left[ (\mu-\xi)\left\{1-\exp\left(\frac{\xi-\mu}{\sigma} \right) \right\}
   - \sigma\left\{\exp\left(\frac{\xi-\mu}{\sigma}\right) \left(\frac{\xi-\mu}{\sigma}-1 \right) +1 \right\} \right]
\nonumber \\
&=& (\mu-\xi) + \tau\exp\left( \frac{\xi-\mu}{\tau} \right) \nonumber \\
&=& |\xi-\mu| + \tau\exp\left( -\frac{|\xi-\mu|}{\tau} \right) \nonumber 
\end{eqnarray}
For $\mu <\xi$
\begin{eqnarray}
 E_L\Bigl[\, |x-\xi|\, \Bigr] 
  &=& \frac{1}{2\sigma} \int_{-\infty}^{\mu} (\xi-x) \exp\left( -\frac{\mu-x}{\sigma} \right)dx
   + \frac{1}{2\sigma} \int_{\mu}^{\xi}( \xi-x) \exp\left( -\frac{x-\mu}{\sigma} \right)dx \nonumber\\
  & & + \frac{1}{2\sigma} \int_{\xi}^{\infty}( x-\xi) \exp\left( -\frac{x-\mu}{\sigma} \right)dx \nonumber\\
  &=& \frac{1}{2\sigma}\left[
  \int_{-\infty}^{0} (\xi-\mu-\sigma y) \exp\left( y \right)dy
 +  \int_{0}^{a}( \xi-\mu-\sigma y) \exp\left( -y \right)dy \right.\nonumber\\
  & & \left. +  \int_{a}^{\infty}( \mu-\xi+\sigma y) \exp\left( -y \right)dy 
\right] \nonumber\\
&=& \frac{1}{2}\biggl[ (\xi-\mu) + \sigma   \nonumber \\
&&   + (\mu-\xi)\left( \exp\left(-\frac{\xi-\mu}{\sigma}\right)-1\right) 
   +\sigma\left\{ \exp\left( -\frac{\xi-\mu}{\sigma}\right)\left( \frac{\xi-\mu}{\sigma}+1\right) - 1 \right\}   \nonumber\\
&& \left. + (\mu-\xi)\exp\left( -\frac{\xi-\mu}{\sigma}\right) 
     + \sigma\left( \frac{\xi-\mu}{\sigma} + 1 \right)\exp\left(-\frac{\xi-\mu}{\sigma} \right) \right] \nonumber \\
&=& (\xi-\mu) + \sigma \exp\left(-\frac{\xi-\mu}{\sigma}\right)  \nonumber 
\end{eqnarray}
Summarizing the above two cases, we have
\begin{eqnarray}
 E_L\Bigl[\, |x-\xi|\, \Bigr] 
  &=& |\xi-\mu| + \sigma \exp\left(-\frac{|\xi-\mu|}{\sigma}\right) .
\end{eqnarray}

\section{Expectation with Respect to the Asymptotic Distribution of the Parameters of the Gauss Distribution Model}

\subsection{True Model:  Gaussian Distribution Model}
In this section, we consider the information criterion for the Laplace distribution model.
We assume that the true model is $g(x|\mu ,\tau )$ and the estimated model is $f(x|\xi ,\sigma )$
defined by:
\begin{eqnarray}
   g(x|\mu ,\tau ) &=& \frac{1}{2\tau } \exp \left( - \frac{|x-\mu |}{\tau} \right) \\
   f(x|\xi ,\sigma ) &=& \frac{1}{2\sigma } \exp \left( - \frac{|x-\xi |}{\sigma} \right).
\end{eqnarray}

Asymptotic distribution of MLE
\begin{eqnarray}
\hat{\sigma} \sim N\left(\left(1-\frac{1}{n}\right)\tau,\frac{\tau^2}{n} \right),  
   \qquad \hat{\sigma} \sim N\left(\left(1-\frac{1}{2n}\right)\tau,\frac{\tau^2}{n} \right)  
  \nonumber 
\end{eqnarray}

\subsection{True Model:  Laplace Distribution Model}
In this subsection, we assume that the true model is a Gauss distribution model $g(x|\mu ,\tau )$ and the estimated model is Laplace distribution model $f(x|\xi ,\sigma )$
defined by:
\begin{eqnarray}
   g(x|\mu ,\tau ) &=& \frac{1}{2\tau } \exp \left( - \frac{|x-\mu |}{\tau} \right) \\
   f(x|\xi ,\sigma^2 ) &=& \frac{1}{\sqrt{2\pi\sigma^2} } \exp \biggl\{ - \frac{(x-\xi)^2}{2\sigma^2} \biggr\}.
\end{eqnarray}
For large $n$, the sampling distributions of the maximum likelihood estimators of $\xi$ and $\tau$ are given by
\begin{eqnarray}
\hat{\xi} \sim N\left(\mu,\frac{2\sigma^2}{n} \right),  
   \qquad \hat{\sigma^2} \sim N\left( \biggl(1-\frac{1}{n}\biggr)\sigma^2,\biggl(1-\frac{2}{n}\biggr)\frac{5\sigma^4}{n} \right)  .
 \nonumber 
\end{eqnarray}
Under this distribution, the expected value of the inverse of $\hat{\sigma}^2$ can be evaluated as
%
\begin{small}
\begin{eqnarray}
E_G\left[ \frac{\sigma^2}{\hat{\sigma}^2}\right]
  &\approx& \tau^2 \int_{-\infty}^{\infty} 
     \frac{1}{\hat{\sigma^2}}\frac{N}{\sqrt{10(N-2)\pi\sigma^4}} \exp\left\{ -\frac{N^2}{10(N-2)\tau^4} 
     \left(\hat{\sigma^2}- \frac{N-1}{N} \tau^2\right)^2 \right\}d\sigma^2  \nonumber \\
  &=& \frac{1}{\sqrt{\pi}} \int_{-\infty}^{\infty} 
      \left\{1 + \frac{1}{N} \left(\sqrt{10(N-2)}y  - 1 \right)\right\}^{-1} 
      \exp\left\{ -y^2 \right\} dy \nonumber  \\
  &=& \frac{1}{\sqrt{\pi}} \int_{-\infty}^{\infty} 
         \biggl\{1 - \frac{cy-1}{N} + \frac{(cy-1)^2}{N^2} - \frac{(cy-1)^3}{N^3} 
                + \frac{(cy-1)^4}{N^4} - \cdots   \biggr\} 
      \exp\left\{ -y^2 \right\} dy \nonumber \\
  &=& \frac{1}{\sqrt{\pi}} \int_{-\infty}^{\infty} 
         \left\{\left(1-\frac{1}{N}\right)^{-1}
         \!\!\!\!- \frac{cy}{N}\left(1 +\frac{2}{N} +\frac{3}{N^2}+ \cdots\right)
         + \frac{c^2y^2}{N^2}\!\!\left(1 +\frac{3}{N} +\frac{6}{N^2} +\cdots \right) 
         + \cdots   \right\} 
      e^{-y^2} dy \nonumber \\
  &=& \frac{1}{\sqrt{\pi}} \left\{ \left(1-\frac{1}{N}\right)^{-1}\!\!\!\!\!\sqrt{\pi}
      + \frac{c^2}{N^2}\left(1 +\frac{3}{N} +\frac{6}{N^2}+ \cdots\right)\!\!\frac{\sqrt{\pi}}{2} 
      + \frac{c^4}{N^4}\left(1 +\frac{5}{N} +\frac{15}{N^2} +\cdots \right)\!\!\frac{3\sqrt{\pi}}{4} 
      + \cdots \right\} \nonumber \\
  &=& \left(1-\frac{1}{N}\right)^{-1}
      + \frac{c^2}{2N^2}\left(1 +\frac{3}{N} +\frac{6}{N^2}+ \cdots\right) 
      + \frac{3c^4}{4N^4}\left(1 +\frac{5}{N} +\frac{15}{N^2} +\cdots \right) 
      + \cdots  \nonumber \\
  &=& \left(1-\frac{1}{N}\right)^{-1}
      + \frac{5(N-2)}{N^2}\left(1 +\frac{3}{N} +\frac{6}{N^2}+ \cdots\right)
      + \frac{15(N-2)^2}{N^4}\left(1 +\frac{5}{N} +\frac{15}{N^2} +\cdots \right) 
      + \cdots  \nonumber \\
  &=& 1 + \frac{6}{N} + \frac{21}{N^2} + \frac{46}{N^3} + O(N^{-4})  .
\label{Eq_Laplace-Gauss_E[1/tau2]}
\end{eqnarray}
\begin{eqnarray}
E_G\left[ (\hat{\mu} - \mu )^2 \right]
  &\approx& \int_{-\infty}^{\infty} (\hat{\mu}-\mu)^2 \sqrt{\frac{N}{2\pi\tau^2}} \exp\left\{ -\frac{N(\hat{\mu}-\mu)^2}{2\tau^2}\right\}d\mu \nonumber \\
  &=& \frac{2\tau^2}{N\sqrt{\pi}} \int_{-\infty}^{\infty} \mu^2  \exp\left\{ -y^2 \right\}dy \nonumber\\
  &=& \frac{2\tau^2}{N\sqrt{\pi}} \frac{\sqrt{\pi}}{2} =\frac{\tau^2}{N} \nonumber 
\end{eqnarray}
\end{small}

Therefore, the bias of the miximized log-likelihood as an estimator of the expected log-likelihood is evaluated as;
%
\begin{eqnarray}
E_G\biggl[ \ell (\hat{\mu},\hat{\tau}^2) &-& N E_G\left[ \log f(X|\hat{\mu} ,\hat{\tau}^2) \right] \biggr]  \nonumber \\
&=& \frac{N}{2} E_L\left[ \frac{2\sigma^2+ (\hat{\xi}-\mu)^2}{\hat{\tau}^2} -1 \right] \nonumber \\
&=& \frac{N}{2} E_L\left[ \frac{\tau^2}{\hat{\tau}^2} \biggl\{ \frac{2\sigma^2+ (\hat{\xi}-\mu)^2}{\hat{\tau}^2} \biggr\} -1 \right] \nonumber \\
  &\approx& \frac{N}{2}E_G\left[ \frac{\tau^2}{\hat{\tau}^2}\right] \left\{1 + \frac{1}{\tau^2} E_G\left[(\mu -\hat{\mu})^2\right] \right\}   -\frac{N}{2} \nonumber \\
  &\approx& \frac{N}{2} \left( 1 + \frac{6}{N} + \frac{21}{N^2} +\frac{46}{N^3} \right) 
            \left( 1 + \frac{1}{N} \right) - \frac{N}{2}  \nonumber \\
  &=& \frac{7}{2} + \frac{27}{2}\frac{1}{N} + \frac{67}{2}\frac{1}{N^2} + O(N^{-3})  .
\label{Eq_Laplace-Gauss_Bias}
\end{eqnarray}

\section{Expectation with Respect to the Asymptotic Distribution of the Parameters of the Laplace Distribution Model}

\subsection{True Model: Laplace Distribution Model}
In this subsection, we assume that the true model is $g(x|\mu ,\tau )$ and the estimated model is $f(x|\xi ,\sigma )$
defined by:
\begin{eqnarray}
   g(x|\mu ,\tau ) &=& \frac{1}{2\tau } \exp \left( - \frac{|x-\mu |}{\tau} \right) \\
   f(x|\xi ,\sigma ) &=& \frac{1}{2\sigma } \exp \left( - \frac{|x-\xi |}{\sigma} \right).
\end{eqnarray}
For large $n$, the sampling distributions of the maximum likelihood estimators of $\xi$ and $\sigma$ are given by
\begin{eqnarray}
\hat{\xi} \sim N\left(\mu,\frac{\tau^2}{N} \right),  
   \qquad \hat{\sigma} \sim N\left( \biggl(1-\frac{2}{N}\biggr)\tau,\frac{\tau^2}{N} \right)  .
 \nonumber 
\end{eqnarray}
Under this distribution, the expected value of the inverse of $\hat{\sigma}$ can be evaluated as
\begin{footnotesize}
\begin{eqnarray}
E\left[\frac{\tau}{\hat{\sigma}}\right] &=&  \int_{-\infty}^{\infty} \frac{\tau}{\hat{\sigma}} \sqrt{\frac{n}{2\pi\tau^2}}
\exp\left\{ -\frac{n}{2\tau^2}\left(\hat{\sigma}-\frac{2n-1}{2n}\tau\right)^2 \right\} d\hat{\sigma}  \nonumber \\
&=&\frac{1}{\sqrt{\pi}} \int_{-\infty}^{\infty} \left\{ 1 + \left(\sqrt{\frac{2}{n}}y - \frac{1}{2n} \right)\right\}^{-1} \exp \left\{ -y^2 \right\} dy \nonumber \\
&\approx& \frac{1}{\sqrt{\pi}} \int_{-\infty}^{\infty} 
   \!\!\left\{ 1 - \left(\sqrt{\frac{2}{n}}y - \frac{1}{2n} \right)
    + \left(\sqrt{\frac{2}{n}}y - \frac{1}{2n} \right)^2 
    - \left(\sqrt{\frac{2}{n}}y - \frac{1}{2n} \right)^3 
    + \left(\sqrt{\frac{2}{n}}y - \frac{1}{2n} \right)^4 + \cdots \right\} 
    \exp \left\{ -y^2 \right\} dy \nonumber \\
&=& \frac{1}{\sqrt{\pi}} \int_{-\infty}^{\infty} 
   \!\!\left\{ \left(1-\frac{1}{2n}\right)^{-1}
   \!\! - \sqrt{\frac{2}{n}}\left(1 - \frac{2}{2n} \right)^{-1}\!\!\!\!y
    + \frac{2}{n}\left(1 - \frac{3}{2n} \right)^{-1}\!\!\!\!y^2 
    - \left(\frac{2}{n}\right)^{\frac{3}{2}}\left(1 - \frac{4}{2n} \right)^{-1}\!\!\!\!y^3 
    + \cdots \right\} 
    \exp \left\{ -y^2 \right\} dy \nonumber \\
&=& \frac{1}{\sqrt{\pi}}  \left\{ \sqrt{\pi}\left(1-\frac{1}{2n}\right)^{-1} + 0 
   +  \frac{2}{n}\frac{\sqrt{\pi}}{2}\left(1-\frac{3}{2n}\right)^{-1} + 0
   +  \frac{4}{n^2}\frac{3\sqrt{\pi}}{4}\left(1-\frac{5}{2n}\right)^{-1} 
   + \cdots \right\} \nonumber \\
&=& \left(1-\frac{1}{2n}\right)^{-1} + \frac{1}{n}\left(1-\frac{3}{2n}\right)^{-1}
    + \frac{3}{n^2}\left(1-\frac{5}{2n}\right)^{-1}
    + \frac{15}{n^2}\left(1-\frac{7}{2n}\right)^{-1}
    + \cdots  \nonumber \\
&=& 1  + \frac{3}{2n} + \frac{19}{4n^2} + \frac{199}{8n^3} +  \cdots 
  \label{Eq_Laplace-Laplace_E[tau/sigma]}
\end{eqnarray}
\end{footnotesize}

Further, the expected value of $\exp\left\{ -\frac{|x-\mu|}{\tau} \right\}$ and 
$|\hat{\xi}-\mu|$ are given by;
%
\begin{eqnarray}
E\left[ \exp\left\{ -\frac{|x-\mu|}{\tau} \right\} \right]  
&=& \int_{-\infty}^{\infty} \exp\left\{ -\frac{|x-\mu|}{\tau} \right\}
\sqrt{\frac{n}{2\pi\tau^2}} \exp\left\{ -\frac{n(x-\mu)^2}{2\tau^2} \right\} dx  \nonumber \\
&=& \frac{1}{\sqrt{\pi}} \int_{-\infty}^{\infty} \exp\left\{ -\frac{\sqrt{2}}{\sqrt{n}} |y| \right\}
   \exp\left\{ -y^2 \right\} dy \nonumber \\
&=& \frac{1}{\sqrt{\pi}} \int_{-\infty}^{\infty} \exp\left\{ - \frac{\sqrt{2}}{\sqrt{n}} |y| 
   -y^2 \right\} dy \nonumber \\
&=& \exp\left( \frac{1}{2n}\right) \textrm{erfc}\left( \frac{1}{\sqrt{2n}} \right) \nonumber \\
&\approx&  1 - \frac{\sqrt{2}}{\sqrt{n\pi}} +\frac{1}{2n} - \frac{\sqrt{2}}{3n\sqrt{n\pi}}
  + \frac{1}{8n^2} + O(n^{-\frac{5}{2}}).  \label{Eq_Laplace-Laplace_E[exp(-|X|]} \\
E\left[ |\hat{\xi}-\mu| \right]
&=& \int_{-\infty}^{\infty} |\hat{\xi}-\mu| \sqrt{\frac{n}{2\pi^2\tau^2}}\exp\left\{ -\frac{n(\hat{\xi}-\mu)^2}{2\pi\tau^2} \right\} d\hat{\xi}  \nonumber \\
 &=& \sqrt{\frac{2}{n}}\tau\int_{-\infty}^{\infty} |y|\exp\left\{-y^2 \right\} dy 
 = \sqrt{\frac{2}{n}}\tau  \label{Eq_Laplace-Laplace_E[|xi-mu|]}   
\end{eqnarray}

Therefore, the bias of the miximized log-likelihood as an estimator of the expected log-likelihood is evaluated as;
\begin{eqnarray}
C_n &=& E \biggl[ \ell(\hat{\xi},\hat{\sigma})
                      -  n E_{L}  \log f(Y|\hat{\xi},\hat{\sigma}) \biggr] \nonumber \\
  &=& n E \left[ \frac{\sigma}{\hat{\tau}}\left\{ \exp\left(-\frac{|\hat{\xi}-\mu|}{\sigma}\right)
       - \frac{|\hat{\xi}-\mu|}{\sigma} \right\} - 1 \right]  \nonumber \\
  &\approx& n E \left[ \frac{\sigma}{\hat{\tau}}\right] 
        \left\{ E\left[\exp\left(-\frac{|\hat{\xi}-\mu|}{\sigma}\right)\right] 
      - \frac{E[\,|\hat{\xi}-\mu|\,]}{\sigma}\right\}  - n   \nonumber \\
  &\approx&  n \left( 1 + \frac{3}{2n} + \frac{19}{4n^2} + \cdots \right)
        \left\{ \exp\left( \frac{1}{2n} \right)\textrm{erfc}\left( \frac{1}{\sqrt{2n}} \right) 
           - \frac{\sqrt{2}}{\sqrt{n}} \right\} - n \nonumber \\
  &\approx& n \left( 1 + \frac{3}{2n} + \frac{19}{4n^2} + \cdots \right)
        \biggl( 1 + \frac{1}{2n} - \frac{\sqrt{2}}{3n\sqrt{n\pi}} + \frac{1}{8n^2}
                  + \cdots \biggr) - n \nonumber\\
  &\approx& 2 - \frac{\sqrt{2}}{3\sqrt{\pi n}} + \frac{45}{8n} .
\end{eqnarray}

\subsection{True Model: Gaussian Distribution Model}

In this subsection, we assume that the true model is $g(x|\mu ,\tau )$ and the estimated model is $f(x|\xi ,\sigma )$
defined by:
\begin{eqnarray}
   g(x|\mu ,\tau ) &=& \frac{1}{\sqrt{2\pi\tau^2}} \exp \left( - \frac{(x-\mu)^2}{2\tau^2} \right) \\
   f(x|\xi ,\sigma ) &=& \frac{1}{2\sigma } \exp \left( - \frac{|x-\xi |}{\sigma} \right).
\end{eqnarray}
For large $n$, the sampling distributions of the maximum likelihood estimators of $\xi$ and $\sigma$ are given by
\begin{eqnarray}
\hat{\xi} \sim N\left(\mu,\frac{\pi\tau^2}{2N} \right),  
   \qquad \hat{\sigma} \sim N\left( \sqrt{\frac{2}{\pi}}\sigma,\frac{\pi-2}{\pi N}\sigma^2 \right) = N\left( \frac{2}{\sqrt{\pi}}\tau,\frac{2(\pi-2)}{\pi N}\tau^2 \right) .
 \nonumber 
\end{eqnarray}
Under this distribution, the expected value of the inverse of $\hat{\sigma}$ can be evaluated as
\begin{eqnarray}
E\left[\frac{\tau}{\hat{\sigma}}\right]
&=& \int_{-\infty}^{\infty}  \frac{\tau}{\hat{\sigma}}  \sqrt{\frac{n}{4(\pi-2)\tau^2}}    \exp\left\{ -\frac{n\pi}{4(\pi-2)\tau^2}\left(\hat{\sigma}- \frac{2}{\sqrt{\pi}}\tau  \right)^2 \right\} d\hat{\sigma}   \nonumber \\
&=& \frac{1}{\sqrt{\pi}} \int_{-\infty}^{\infty} \frac{\sqrt{\pi}}{2}
     \left( 1 + \sqrt{\frac{\pi-2}{n}}y \right)^{-1}\exp \left\{ -y^2 \right\} dy \nonumber \\
&\approx& \frac{1}{2} \int_{-\infty}^{\infty} 
   \left\{ 1 - \sqrt{\frac{\pi-2}{n}}y + \frac{\pi-2}{n}y^2 - \left(\frac{\pi-2}{n}\right)^{3/2}\!\!y^3 + \frac{(\pi-2)^2}{n^2}y^4 + \cdots \right\} 
    \exp \left\{ -y^2 \right\} dy \nonumber \\
&=& \frac{1}{2}  \left\{ \sqrt{\pi} + 0 +  \frac{\pi-2}{n}\frac{\sqrt{\pi}}{2} + 0
   +  \frac{(\pi-2)^2}{n^2}\frac{3\sqrt{\pi}}{4} + 0 + \frac{(\pi-2)^3}{n^3}\frac{15\sqrt{\pi}}{8}  + \cdots \right\} \nonumber \\
&=& \frac{\sqrt{\pi}}{2} \left\{ 1 + \frac{\pi-2}{2n} + \frac{3(\pi-2)^2}{4n^2} + \frac{15(\pi-2)^3}{8n^3} + \cdots  \right\}   \label{Eq_Gauss-Laplace_E[tau-sigma]}
\end{eqnarray}

Further, the expected value of $\exp\left\{ -\frac{|x-\mu|}{\tau} \right\}$ and 
$\frac{\hat{\xi}-\mu}{\sqrt{2}\sigma}\textrm{erf}\left(\frac{\hat{\xi}-\mu}{\sqrt{2}\sigma} \right)$ are given by;
%
%
\begin{eqnarray}
E\left[ \exp\left\{ -\frac{(\hat{\xi}-\mu)^2}{2\sigma^2} \right\} \right]  
&=& \int_{-\infty}^{\infty} \exp\left\{ -\frac{(\hat{\xi}-\mu)^2}{2\sigma^2} \right\}
\sqrt{\frac{n}{\pi^2\sigma^2}} \exp\left\{ -\frac{n(\hat{\xi}-\mu)^2}{\pi\sigma^2} \right\} d\hat{\xi}  \nonumber \\
&=& \sqrt{\frac{n}{\pi^2\sigma^2}} \int_{-\infty}^{\infty} \exp\left\{ -\frac{(x-\mu)^2}{2\sigma^2} - \frac{n(x-\mu)^2}{\pi\sigma^2} \right\}
    dx \nonumber \\
&=& \frac{\sqrt{n}}{\pi\sigma} \int_{-\infty}^{\infty} \exp\left\{ - \frac{\pi+2n}{2\pi\sigma^2}(x-\mu)^2 \right\} dx \nonumber \\
&=& \frac{\sqrt{n}}{\pi\sigma} \frac{\sqrt{2\pi}\sigma}{\sqrt{\pi+2n}} \int_{-\infty}^{\infty} \exp\left\{ -y^2 \right\} dy \nonumber \\
&=&  \frac{\sqrt{2n}}{\sqrt{\pi(\pi+2n)}}\sqrt{\pi}  = \frac{\sqrt{2n}}{\sqrt{2n+\pi}} \label{Eq_Gauss-Laplace_E[exp(-|X|)]} 
\end{eqnarray}
%
\begin{eqnarray}
E\left[ \frac{\hat{\xi}-\mu}{\sqrt{2}\sigma} \textrm{erf}\left( \frac{\hat{\xi}-\mu}{\sqrt{2}\sigma} \right) \right]
&=& \int_{-\infty}^{\infty} \frac{\hat{\xi}-\mu}{\sqrt{2}\sigma}
       \textrm{erf}\left( \frac{\hat{\xi}-\mu}{\sqrt{2}\sigma} \right) 
       \sqrt{\frac{n}{\pi^2\sigma^2}}\exp\left\{ -\frac{n(\hat{\xi}-\mu)^2}{\pi\sigma^2} \right\} 
    d\hat{\xi}  \nonumber \\
&=& \frac{1}{\sqrt{\pi}}  \int_{-\infty}^{\infty} \left(\sqrt{\frac{\pi}{2n}}y \right)
       \textrm{erf}\left(\sqrt{\frac{\pi}{2n}}y \right) 
       \exp\left\{ -y^2 \right\} dy  \nonumber \\
 &=& \frac{1}{\sqrt{\pi}}\frac{\frac{\pi}{2n}}{\sqrt{\frac{\pi}{2n}+1}}
 = \frac{\sqrt{\pi}}{\sqrt{n(2n+\pi)}}  \label{Eq_Gauss-Laplace_E[|xi-mu|]} \\
\int_{-\infty}^{\infty} \textrm{erf}(ay) \,y\, e^{-y^2} dy &=& \frac{a}{\sqrt{a^2+1}} \nonumber\\
\sqrt{\frac{2n}{2n+\pi}} &=& 1 - \frac{\pi}{4n} + \frac{3\pi^2}{32n^2} - \frac{5\pi^3}{128n^3} + O(n^{-4}) \nonumber \\
\frac{\pi}{\sqrt{n(2n+\pi)}} &=&  \frac{1}{\sqrt{2}}\left( \frac{\pi}{n} 
    - \frac{\pi^2}{4n^2} + \frac{3\pi^3}{32n^3} \right) + O(n^{-4}) \nonumber \\
  &=& \frac{\sqrt{2}\pi}{2n} 
    - \frac{\sqrt{2}\pi^2}{8n^2} + \frac{3\sqrt{2}\pi^3}{64n^3} + O(n^{-4}) \nonumber \\
\sqrt{\frac{2n}{2n+\pi}} + \!\frac{\sqrt{2}\pi}{\sqrt{n(2n+\pi)}} 
   &=&  \!1 + \frac{(2\sqrt{2}-\!1)\pi}{4n} - \frac{(4\sqrt{2}-\!3)\pi^2}{32n^2} + \frac{(6\sqrt{2}-\!5)\pi^3}{128n^3} + O(n^{-4}) \label{Eq_Gauss-Laplace_sqrt(2n/2n+pi)}  
\end{eqnarray}
%
Therefore, the bias of the miximized log-likelihood as an estimator of the expected log-likelihood is evaluated as;
\begin{small}\begin{eqnarray}
\lefteqn{ E_{\hat{G}}\left[ \ell (\hat{\xi},\hat{\sigma} ) - E_L[\log f(x|\hat{\xi},\hat{\sigma} )] \right] }
\nonumber \\
 &=& E_{\hat{G}}\left[ \frac{1}{\hat{\sigma}}\sqrt{\frac{2\sigma^2}{\pi}}\exp\left\{-\frac{(\hat{\xi}-\mu)^2}{2\sigma^2}\right\}
      +  \frac{\hat{\xi}-\mu}{\hat{\sigma}}\textrm{erf}\left( \frac{\hat{\xi}-\mu}{\sqrt{2\sigma^2}} \right)  - 1 \right]   \nonumber \\
 &=&  E_{\hat{G}}\left[ \frac{\tau}{\hat{\sigma}}\left\{ \frac{2}{\sqrt{\pi}}\exp\left\{-\frac{(\hat{\xi}-\mu)^2}{2\sigma^2}\right\}
      +  \frac{\hat{\xi}-\mu}{\tau}\textrm{erf}\left( \frac{\hat{\xi}-\mu}{\sqrt{2}\sigma} \right) \right\}  - 1 \right]  \nonumber \\
 &\approx& E_{\hat{G}}\left[ \frac{\tau}{\hat{\sigma}}\right] 
    \left( \frac{2}{\sqrt{\pi}} E_{\hat{G}}\left[ \exp\left\{-\frac{(\hat{\xi}-\mu)^2}{2\sigma^2}\right\} \right]
  +  2 E_{\hat{G}}\left[ \frac{\hat{\xi}-\mu}{\sqrt{2}\sigma}\textrm{erf}\left( \frac{\hat{\xi}-\mu}{\sqrt{2}\sigma} \right) \right] \right) 
            - 1   \nonumber \\
 &\approx& \frac{\sqrt{\pi}}{2} \left( 1 + \frac{(\pi-2)}{2n} + \frac{3(\pi-2)^2}{4n^2} + \frac{15(\pi-2)^3}{8n^3} \right)
      \left\{ \frac{2}{\sqrt{\pi}} \frac{\sqrt{2n}}{\sqrt{2n+\pi}} 
      +  \frac{2\sqrt{\pi}}{\sqrt{n(2n+\pi)}}    \right\} - 1 \nonumber \\
 &=&  \left( 1 + \frac{(\pi-2)}{2n} + \frac{3(\pi-2)^2}{4n^2} + \frac{15(\pi-2)^3}{8n^3} \right)
    \left\{ \frac{\sqrt{2n}}{\sqrt{2n+\pi}} 
      + \frac{\pi}{\sqrt{n(2n+\pi)}}   \right\} - 1 \nonumber \\
 &\approx& \left\{ 1 + \frac{1}{n}\!\!\left( \frac{\pi}{2}-\!1\right) + \frac{1}{n^2}\!\!\left(\frac{3}{4}\pi^2 \!-\!3\pi + 3\right) + \frac{1}{n^3}\!\!\left(\frac{15}{8}\pi^3 \!-\! \frac{45}{4}\pi^2 \!+\! \frac{45}{2}\pi \!-\! 15\right) \right\} \nonumber \\
&& \qquad  \times  \biggl\{ 1 + \frac{(2\sqrt{2}-1)\pi}{4n} - \frac{(4\sqrt{2}-3)\pi^2}{32n^2}
     + \frac{(6\sqrt{2}-5)\pi^3}{128n^3}   \biggr\}  - 1   \nonumber \\
&=& \frac{1}{n} \left( \frac{2\sqrt{2}+1}{4}\pi-1\right)
  + \frac{1}{n^2}\left(\frac{4\sqrt{2}+31}{32}\pi^2 -\frac{2\sqrt{2}+13}{4}\pi +3\right)\nonumber \\
&& \qquad   + \frac{1}{n^3}\left( \frac{46\sqrt{2}+217}{128}\pi^3 - \frac{44\sqrt{2}+339}{32}\pi^2
       + \frac{6\sqrt{2}+87}{4}\pi - 15 \right).
\end{eqnarray}\end{small}

%
%

\section{Supplementary Numerical Results}

\subsection{Effect of Variance Reduction Method}

\begin{table}[h]
\caption{True Bias Obtained by Monte Carlo Evaluation. Number of iterations in Monte Carlo evaluation, $NS=10^8$ for $N$=25 and $10^7$ for $N$=400 and $NS$=$10^6$ for $N$=1,600. G and L denote the Gauss model and Laplace model, respectively.}
\label{Tab_Decomposition_of_C}
\begin{center}
\begin{tabular}{cc|c|cccc|rrrr}
 & &       &\multicolumn{4}{|c|}{Mean} &\multicolumn{4}{|c}{Variance} \\
\cline{4-11}
D & M &$N$ &  25  & 100  & 400 & 1,600 &  25\hspace{3mm}  & 100\hspace{3mm} & 400\hspace{3mm} & 1,600 \\
\hline
 & &$C_1$  & 1.038 & 1.009 & 1.003 & 1.001 & 1.078 &  1.020 &   1.007 &   1.009 \\
 & &$C_2$  &-0.001 &-0.001 &-0.006 & 0.005 &12.501 & 50.004 & 200.049 & 800.385 \\
G&G&$C_3$  & 1.234 & 1.053 & 1.013 & 1.004 & 2.536 &  1.280 &   1.068 &   1.025 \\
 & &$C$    & 2.272 & 2.062 & 2.010 & 2.010 &22.526 & 57.613 & 207.238 & 808.216 \\
 & &$C_1+C3$&2.273 & 2.062 & 2.016 & 2.005 & 6.622 &  4.516 &   4.131 &   4.064 \\
\hline
 & &$C_1$  & 1.085 & 1.029 & 1.014 & 1.007 & 1.100 &  1.038 &   1.016 &   1.021 \\
 & &$C_2$  &-0.001 & 0.001 &-0.004 & 0.014 &25.000 & 99.974 & 399.901 &1600.689 \\
L&L&$C_3$  & 1.189 & 1.067 & 1.030 & 1.015 & 1.744 &  1.236 &   1.098 &   1.060 \\
 & &$C$    & 2.274 & 2.096 & 2.040 & 2.037 &32.627 &106.473 & 406.097 &1611.915 \\
 & &$C_1+C3$&2.275 & 2.096 & 2.044 & 2.022 & 5.365 &  4.431 &   4.172 &   4.134 \\
\hline
 & &$C_1$  & 1.702 & 1.732 & 1.745 & 1.750 & 3.384 &  3.582 &   3.625 &   3.643 \\
 & &$C_2$  &-0.001 & 0.002 &-0.001 & 0.023 &31.240 &124.939 & 500.070 &2000.652 \\
L&G&$C_3$  & 2.173 & 1.840 & 1.771 & 1.757 & 9.774 &  4.735 &   3.888 &   3.723 \\
 & &$C$    & 3.874 & 3.574 & 3.515 & 3.530 &59.593 &145.058 & 518.401 &2021.725 \\
 & &$C_1+C3$&3.875 & 3.572 & 3.516 & 3.507 &22.140 & 15.887 &  14.850 &  14.689 \\

\hline
 & &$C_1$  & 1.112 & 1.073 & 1.072 & 1.071 & 1.264 &  1.301 &   1.340 &   1.385 \\
 & &$C_2$  &-0.000 &-0.000 &-0.004 & 0.011 &14.271 & 57.083 & 228.420 & 913.874 \\
G&L&$C_3$  & 1.143 & 1.079 & 1.073 & 1.072 & 1.591 &  1.408 &   1.398 &   1.414 \\
 & &$C$    & 2.254 & 2.152 & 2.141 & 2.154 &19.657 & 61.897 & 233.162 & 919.263 \\
 & &$C_1+C3$&2.255 & 2.153 & 2.145 & 2.143 & 5.408 &  5.244 &   5.367 &   5.539 \\
\hline
\end{tabular}
\end{center}
\end{table}

The expected value of $C_2$ is zero, but its variance increases in proportion to the number of data. Therefore, as the number of data $N$ increases, most of the error in $C$ is due to $C_2$. 
Using this property, calculating $C_1 + C_3$ instead of directly evaluating $C$ yields a much more accurate estimate than $C$. 
This variance reduction method is particularly useful in the calculation of the bootstrap estimate $B_n$.

\end{document}